\documentclass[useAMS,usenatbib]{mn2e}
\usepackage{graphicx}
\usepackage[authoryear]{natbib}
\usepackage{longtable}
\usepackage[caption=false]{subfig}

\DeclareCaptionLabelFormat{continued}{#1~#2 (Continued)}
\title[H$\alpha$ kinematics of compact groups]{The H$\alpha$ kinematics of interacting galaxies in 12 compact groups\thanks{Based on observations made with ESO Telescopes at the La Silla Observatory and the Observatoire de Haute Provence, France. Based on observations obtained at the Gemini Observatory, which is operated
by the Association of Universities for Research in Astronomy, Inc.,
under a cooperative agreement with the NSF on behalf of the Gemini partnership:
the National Science Foundation (United States), the Science and
Technology Facilities Council (United Kingdom), the National Research
Council (Canada), CONICYT (Chile), the Australian Research Council
(Australia), Minist\'erio da Ci\^encia e Tecnologia (Brazil) and Ministerio de
Ciencia, Tecnolog\'ia e Innovaci\'on Productiva (Argentina) -- Observing run:
GS-2013B-Q-27.}}

\author[S. Torres-Flores et al.]
{
\parbox[t]{\textwidth} {S. Torres-Flores$^{1}$\thanks{E-mail: storres@dfuls.cl}, P. Amram$^{2}$, C. Mendes de Oliveira$^{3}$, H. Plana$^{4}$, C. Balkowski$^{5}$, M. Marcelin$^{2}$, \& D. Olave-Rojas$^{1}$}
\vspace*{6pt}\\
$^1$Departamento de F\'isica, Universidad de La Serena, Av. Cisternas 1200 Norte, La Serena, Chile \\
$^2$Aix-Marseille Universit\'e, CNRS, LAM (Laboratoire d'Astrophysique de Marseille), 13388, Marseille, France \\
$^3$Departamento de Astronomia, Instituto de Astronomia, Geof\'isica e Ci\^encias Atmosf\'ericas da USP, S\~ao Paulo, Brazil \\
$^4$Laborat\'orio de Astrof\'isica Te\'orica e Observacional, Universidade Estadual de Santa Cruz, Ilh\'eus, Brazil\\
$^5$GEPI, Observatoire de Paris, Paris University Denis Diderot \& CNRS, 5 place Jules Janssen, Meudon, France
}

\begin{document}
\maketitle

\begin{abstract}

We present new Fabry-Perot observations for a sample of 42 galaxies located in twelve compact groups of galaxies: HCG 1, HCG 14, HCG 25, HCG 44, HCG 53, HCG 57, HCG 61, HCG 69, HCG 93, VV 304, LGG 455 and Arp 314. From the 42 observed galaxies, a total of 26 objects are spiral galaxies, which range from Sa to Im morphological types. The remaining 16 objects are E, S0 and S0a galaxies. Using these observations, we have derived velocity maps, monochromatic and velocity dispersion maps for 24 galaxies, where 18 are spiral, three are S0a, two are S0 and one is an Im galaxy. From the 24 velocity fields obtained, we could derive rotation curves for 15 galaxies; only two of them exhibit rotation curves without any clear signature of interactions. Based on kinematic information, we have evaluated the evolutionary stage of the different groups of the current sample. We identify groups that range from having no H$\alpha$ emission to displaying an extremely complex kinematics, where their members display strongly perturbed velocity fields and rotation curves. In the case of galaxies with no H$\alpha$ emission, we suggest that past galaxy interactions removed their gaseous components, thereby quenching their star formation. However, we can not discard that the lack of H$\alpha$ emission is linked with the detection limit for some of our observations.

\end{abstract}


\section{Introduction}

Nearby compact groups provide ideal laboratories for studying the
effects of ongoing collisions on the structure and dynamics of galaxies.
Given their proximity, a detailed study of these systems can help us 
understand in detail some interaction effects that are common in the
distant Universe, where the merger rate was higher than today (e.g.
L\'opez-Sanjuan et al. 2013).

During the last decade, H$\alpha$ Fabry-Perot observations of nearby compact groups were used to derive the spatial distribution and
kinematics of the warm gas content of galaxies in these systems (e. g. Mendes de Oliveira et al. 1998). The
warm gas is a tracer of the potential in a galaxy and a detailed
analysis of the kinematics of the emission-line velocity field can be used to determine the influence of the galaxy interactions in the evolution of compact group galaxies, when compared with galaxies in other environments (e. g. using the Tully-Fisher relation, Torres-Flores et al. 2013).

\begin{figure*}
\centering
\includegraphics[width=0.9\textwidth]{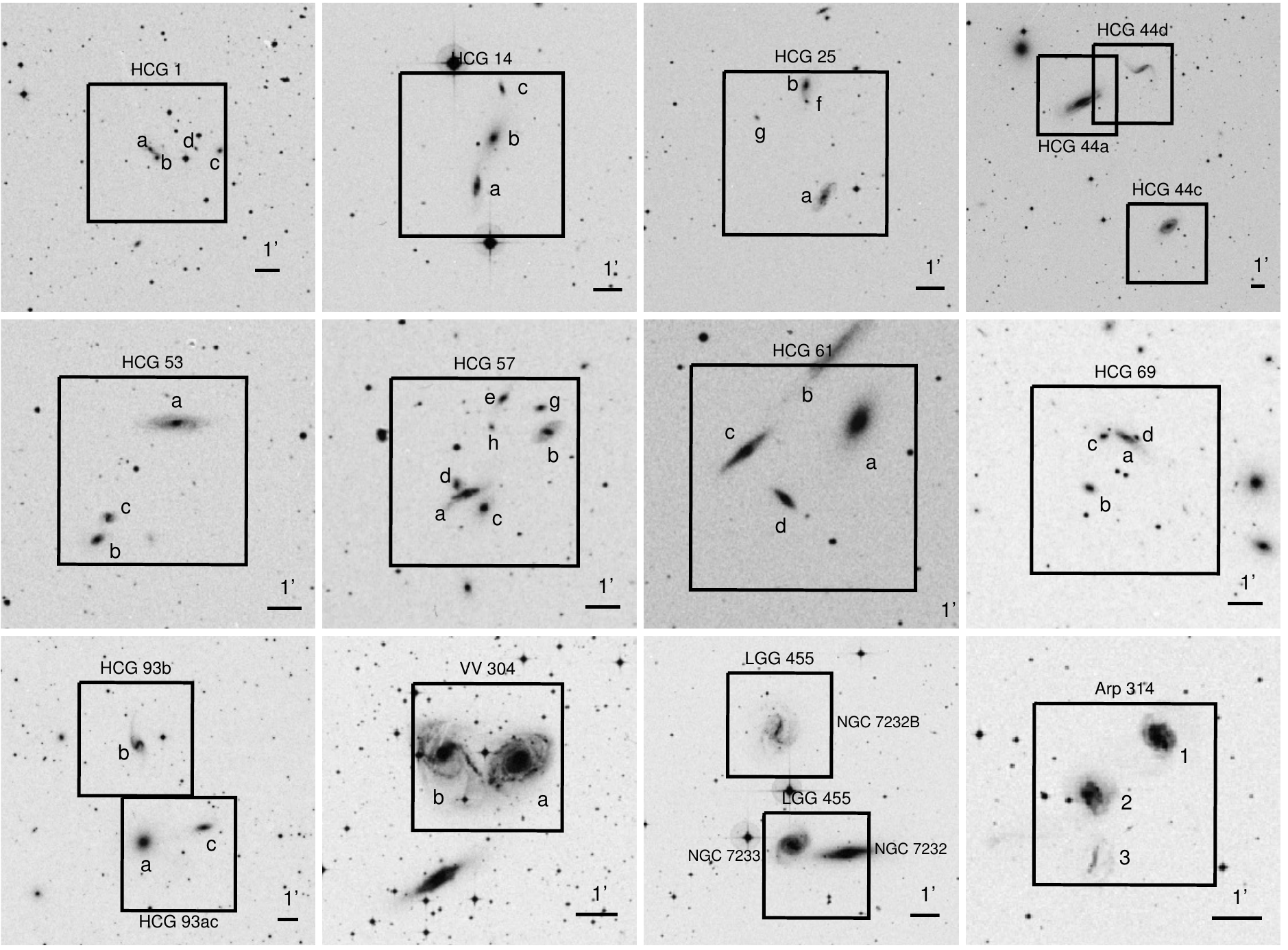}
\caption{Optical DSS images of the set of twelve compact groups analysed in this paper. In all panels (where North is at the top and East is to the left), black rectangles indicate the field-of-view of the
instrument, in which we derived the H$\alpha$ monochromatic images,
velocity fields and dispersion maps by using the Fabry-Perot data cubes.
The horizontal line at the bottom right of each panel indicates a scale
of 1'.}
\label{dss}
\end{figure*}

Successive mass accretion marks the history of galaxies in these
compact interacting systems. The study of the observed kinematics
of a system retains a memory of the accretion process, which drives
galaxy evolution. Misaligned stellar and gas major axes position
angles, anomalous kinematics structures, complex H$\alpha$ distribution,
distorted spiral arms and disagreement between both sides of the
rotation curves are a few of the most common indicators that galaxies
have experienced or are experiencing a collision. All of these have
previously been identified in compact groups (see Amram et al. 2003
and references in that paper). Also, the formation of tidal tails in compact group galaxies is a typical signature of galaxy-galaxy collision. For example, Renaud et al. (2010) performed N-body simulations of the compact group Stephan' Quintet, recovering on this way the main tails visible in this strongly interacting system.

It is crucial to enlarge the set of interacting and merging candidate galaxies for which we have measured velocity fields in order to shed light on their formation/evolution processes. The present study adds twelve new compact groups (and velocity fields for 24 galaxies) to a total of already 25 groups (with velocity fields derived for 58 galaxies\footnote{HCG FabryÐPerot data are available at http://fabryperot.oamp.fr}) already published in the last 15 years  (Mendes de Oliveira et al. 1998, Plana et al. 2003, Amram et al. 2003, Torres-Flores et al. 2009, 2010, 2013) using Fabry-Perot observations. Besides these works, other kinematical studies of compact groups published during the last decade used HI data (e. g. Verdes-Montenegro et al. 2005, Borthakur et al. 2010). In addition, Nishiura et al. (2000) used longslit observations to analyse the kinematics of HCG galaxies and recently Vogt et al. (2013) used integral field spectroscopy to study the galactic winds in the group HCG 16.

The organization of this article is as follows: Section 2 gives details
of the observations and data reduction. In Section 3, we present the
results for the internal kinematics and the mass determinations of the
individual galaxies of the groups. Section 4 contains the discussion and
conclusions.

\section{Observation and data reduction}

\subsection{The Sample}

We have obtained new Fabry-Perot observations for 42 galaxies in
twelve small groups: HCG 1, HCG 14, HCG 25, HCG 44, HCG 53,
HCG 57, HCG 61, HCG 69, HCG 93, VV 304, LGG 455 and LGG 467. Most
of the targets selected here were part of a larger survey study of
the kinematics of galaxies in Hickson Compact Groups (e.g. Torres-Flores
et al. 2010). To the Hickson sample we have added other three systems which
are not included in the Hickson (1992) catalogue. These are LGG 455
and LGG 467 (hereafter LGG 467 will be named Arp 314), listed in the catalogue of compact
groups of Garcia (1995), and in addition, the well known galaxy triplet
VV 304, which is the compact core of a looser group. VV 304 has a median
separation between the two brightest galaxies lower than 3 kpc and the
centre of the next bright neighbour at the same redshift is about
40 kpc of the centre of the triplet, which makes this system similar
to other compact configurations from the Hickson et al. (1992) and
Garcia (1995) catalogues. Most of the systems studied here present
clear signatures of interactions, making them ideal targets to study
the kinematics of the warm ionized gas aiming at studying galaxy evolution in dense environments. All the
groups listed above are located in the nearby universe, having
radial velocities ranging from 1200 km s$^{-1}$ to $\sim$10000 km s$^{-1}$.
In Fig. \ref{dss} we show a DSS optical image of each target and in Table \ref{mainproperties} we list the main properties of each observed galaxy.

\begin{table*}
\begin{minipage}[h!]{\textwidth}
\caption{Main properties of the sample}
\label{mainproperties}
\centering
\renewcommand{\footnoterule}{}  % to avoid a line before footnotes
\begin{tabular}{lcccccccccc}
\hline \hline
(1)\footnotetext{(1) Galaxy ID; (2) Most commonly used name for each galaxy, when available; (3) Right Ascension; (4) Declination;
(5) Morphological Type; (6) T-type morphological type; (7) Systemic velocity; (8) Distance in Mpc (corrected for the Virgo, Great Attractor and Shapley supercluster infall. Values were taken from NED); (9) B-band magnitude; (10) Optical (B-R) colors; (11) Seeing (arcsec). In the case of HCG galaxies,  all quantities were taken from Hickson
1993, except the Right Ascension and Declination that were taken from
the Vizier database and the T-type morphological classification, which was taken from Hickson et al. (1989). For other galaxies, the information was taken from
the NED database, except for the morphological types that were taken from the HyperLeda database.}&(2)&(3)&(4)&(5)&(6)& (7) & (8)& (9) & (10)& (11)\\
ID & ID & $\alpha$ (2000) & $\delta$ (2000) & Morph & T &V$_{syst}$ & Distance &B$_{Tc}$ & B-R & Seeing\\
      & (Alternative name)   &    h:m:s  & d:m:s   &   type  & type &km s$^{-1}$ & Mpc & mag & mag & Arcsec\\
\hline
HCG 1a &  UGC 00248 NED01  & 00:26:07.14 & +25:43:30.2& Sc  &    7 & 10237   & 137.8 &14.43 & 1.29 & 4.5 \\
HCG 1b &  UGC 00248 NED02  & 00:26:05.99 & +25:43:08.2& Im  &   12 & 10266  & 138.2 &15.04 & 1.36 & 4.5 \\
HCG 1c &  ...  & 00:25:54.48 & +25:43:24.0& E0  &    0 & 10056   & 135.3 &15.57 & 1.69 & 4.5 \\
HCG 1d  &  ...   & 00:25:58.86 & +25:43:29.7& S0  &    1 & 10120  & 136.2 &16.50 & 1.60 & 4.5 \\
HCG 14a & MCG -01-06-022   & 01:59:49.82 & -07:03:32.9& Sb  &    5 & 5929    & 77.9 &14.77  & 1.86 & 3.6\\
HCG 14b & MCG -01-06-020   & 01:59:52.23 & -07:05:11.5& E5  &    0 & 5365    & 71.3  &14.17  & 1.54 & 3.6\\
HCG 14c & MCG -01-06-019   & 01:59:48.69 & -07:01:50.9& Sbc &    6 & 5145   & 68.3 &16.58  & 1.94 & 3.6 \\
HCG 25a & UGC 02690   & 03:20:43.15 & -01:06:33.7& SBc &    7 & 6285  & 83.6  &13.86  & 1.05 & 4.4\\
HCG 25b & UGC 02691 NED01     & 03:20:45.59 & -01:02:41.5& SBa &    3 & 6408  & 84.2 &14.45  & 1.75 & 4.4\\
HCG 25f  & UGC 02691 NED02      & 03:20:45.55 & -01:03:15.3& S0  &    1 & 6279     & 83.0  &16.98  & 1.52 & 4.4\\
HCG 25g &  ...   & 03:20:52.32 & -01:03:48.5& S0  &    1 & 12179  & 164.0 &16.93  & 1.70 & 4.4\\
HCG 44a  &  NGC 3190   & 10:18:06.12 & +21:49:52.8& Sa  &    3 & 1293   & 22.6 &11.52  & 1.28 & 2.3 \\
HCG 44c  & NGC 3185   & 10:17:38.73 & +21:41:17.6& SBc &    7 & 1218  & 21.6 &12.55  & 0.97 & 2.3 \\
HCG 44d  & NGC 3187     & 10:17:48.00 & +21:52:23.9& Sd  &    9 & 1579   & 27.3  &13.09  & 0.26 & 2.2 \\
HCG 53a &  NGC 3697  & 11:28:50.37 & +20:47:43.2& SBbc&    6 & 6261& 92.9  &12.91   & 1.30 & 2.7 \\
HCG 53b &  NGC 3697C   & 11:29:00.00 & +20:44:22.0& S0  &    1 & 6166  & 93.9  &14.73   & 1.57 & 2.7\\
HCG 53c &  NGC 3697B   & 11:28:58.55 & +20:45:00.1& SBd &    9 & 6060 & 92.7  &14.81   & 1.00 & 2.7 \\
HCG 57a &  NGC 3753    & 11:37:53.78 & +21:58:51.0& Sb  &    5 & 8727  & 126.5  &13.99  & 1.73 & 5.9\\
HCG 57b &  NGC 3746    & 11:37:43.68 & +22:00:33.6& SBb &    5 & 9022& 130.5  & 14.32  & 1.55 & 5.9\\
HCG 57c  &  NGC 3750   & 11:37:51.75 & +21:58:25.8& E3  &    0 & 9081  & 131.3  &14.63  & 1.57 & 5.9\\
HCG 57d & NGC 3754     & 11:37:55.12 & +21:59:08.3& SBc &    7 & 8977 & 129.9 &14.51  & 1.26 & 5.9\\
HCG 57e & NGC 3748     & 11:37:49.17 & +22:01:32.8& S0a &    2 & 8992 & 127.7 &15.37  & 1.71 & 5.9\\
HCG 57g & NGC 3745     & 11:37:44.61 & +22:01:15.0& SB0 &    1 & 9416 & 136.5 &15.84  & 1.65 & 5.9\\
HCG 57h &  ...    & 11:37:50.69 & +22:00:42.7& SBb &    5 &           &  133.5 &16.75  & 1.53 & 5.9\\
HCG 61a &  NGC 4169   & 12:12:18.55 & +29:10:47.3& S0a &    2 & 3784 & 59.3  &12.82  & 1.50 & 2.8\\
HCG 61c &   NGC 4175   & 12:12:30.95 & +29:10:06.7& Sbc &    6 & 3956 &  62.4 &13.53   & 1.56 & 2.8\\
HCG 61d &  NGC 4174   & 12:12:26.77 & +29:08:56.8& S0  &    1 & 3980 &  62.9  &14.12   & 1.40 & 2.8\\
HCG 69a &  UGC 08842 NED02     & 13:55:29.86 & +25:04:25.9& Sc  &    7 & 8856 &  129.5  &14.94   & 1.57 & 2.3\\
HCG 69b &   ...   & 13:55:34.41 & +25:02:58.4& SBb &    5 & 8707 & 127.7  &15.59   & 1.49 & 2.3\\
HCG 69c & UGC 08842 NOTES01    & 13:55:32.62 & +25:04:27.4& S0  &    1 & 8546 &  125.3    &14.94   & 1.45 & 2.3\\
HCG 69d &  UGC 08842 NED01    & 13:55:28.35 & +25:04:24.7& SB0 &    1 & 9149&  129.6 &16.06   & 1.51 & 2.3\\
HCG 93a &  NGC 7550    & 23:15:16.03 & +18:57:41.3& E1  &    0 & 5140 &  69.2  &12.61   & 1.60 & 2.8\\
HCG 93b &  NGC 7549    & 23:15:17.21 & +19:02:29.7& SBd &    9 & 4672 &  64.8  &13.18   & 1.30 & 3.6\\
HCG 93c &  NGC 7547   & 23:15:03.64 & +18:58:23.2& SBa &    3 & 5132 &  65.2  &13.94   & 1.79 & 2.8\\
VV 304a  &  NGC 6769    & 19:18:22.70 & -60:30:04.0 & SABb & ... & 3686 & 54.0   &12.55 & ... & 1.5 \\
VV 304b  &  NGC 6770    & 19:18:37.30 & -60:29:47.0 & Sb  & ... & 3841 &   56.4 &12.83 & ... & 1.5 \\
NGC 7232 & ...   & 22:15:38.00 & -45:51:00.0 & S0a & ... & 1915 &    27.3 &12.95 & ... & 1.2\\
NGC 7232B& ...  & 22:15:52.40 & -45:46:50.0 & SBm & ... & 2160 &   30.9  & 13.9 & ... & 1.3 \\
NGC 7233  & ...   & 22:15:49.00 & -45:50:47.0 & S0a & ... & 1841 &  26.3   &13.09 & ... & 1.2 \\
Arp 314-1 & MCG -01-58-009    & 22:58:02.20 & -03:46:11.0 & Sbc & ... & 3687 &   51.1  &13.70 & ...& 1.1 \\
Arp 314-2 & MCG -01-58-010   & 22:58:07.90 & -03:47:20.0 & SBc  & ... & 3731 &   51.7  &13.81 &... & 1.1\\
Arp 314-3 & MCG -01-58-011    & 22:58:07.30 & -03:48:38.0 & SBd  & ... & 3701 &  51.3  &16.00 & ... & 1.1 \\
\hline
\end{tabular}
\end{minipage}
\end{table*}

\begin{table*}
\centering
\caption{Journal of Fabry-Perot observations \label{observinglog}}
\begin{tabular}{llll}
\hline
Observations & Telescope & ESO 3.6m & OHP 1.93m  \\
             & Equipment @ Cassegrain focus  &  CIGALE & GHASP  \\
             & Date & 2000, Sept.  & 2011, April \&   2012, October  \\
Spatial sampling  & Total Field & 207"$\times $207''  & 353''$\times
$353''\\
                  & Pixel Size (binned) & 0.405'' & 0.69'' \\
Calibration & Neon Comparison light  & $\lambda$ 6598.95 \AA &   $\lambda$ 6598.95 \AA  \\
Fabry--Perot & Interference Order & 793 @ $\lambda$ 6562.78 \AA & 798 @ $\lambda$ 6562.78 \AA\\
         & Free Spectral Range at H$\alpha$  &378~km~s$^{-1}$ (8.28 \AA)& 376~km~s$^{-1}$ (8.22 \AA) \\
         & Finesse at H$\alpha$  &11.4  &11.9 (run 2011) --11.0 (run 2012)  \\
         & Instrumental FWHM &  33~km~s$^{-1}$ &  32~km~s$^{-1}$ (run 2011) -- 34~km~s$^{-1}$ (run 2012) \\
         & Spectral resolution at H$\alpha$  & 12682 & 12774 \\
Spectral sampling & Number of Scanning Steps  &32 & 32 \\
     & Sampling Step  & 0.26 \AA\ (11.81~km~s$^{-1}$) & 0.26 \AA\ (11.75~km~s$^{-1}$) \\
Detector & IPCS & GaAs Tube & GaAs Tube \\
\hline
\end{tabular}
\end{table*}

\subsection{Observations}
\label{observations}

The observations of HCG 1, 14, 25 and 93 were carried out in October 2012, using the GHASP Fabry-Perot instrument mounted on the 1.93 m
telescope at the Observatoire de Haute Provence (OHP). The groups HCG 44, 53, 57, 61 and
69 were observed with the same instrument and observing setup, in
April 2011. For HCG 14 we have observed two fields, which cover the
members HCG 14a, b and c.  For HCG 57 and HCG 69 we have observed
two fields, which cover members a, b, c, d, e, g, h and a, b ,c , d, respectively. Systems VV 304, LGG 455 and Arp 314 were observed
by using the Fabry-Perot instrument CIGALE mounted on the European
Southern Observatory (ESO) 3.6 m telescope at La Silla (Chile)
in September 2000.

For the ESO and OHP observations, the interference order of the Fabry-Perot was $p=793$ at H$\alpha$ and $p=798$ at H$\alpha$, respectively, where
the free spectral range (FSR) was 378 km s$^{-1}$ and 376 km s$^{-1}$ in each case. For both observations we scanned 32 steps, which gave us sampling steps of $\sim$11.8 km s$^{-1}$ in both cases (R$\sim$12000). The information was recorded by
using a photon counting system, where the pixel size was
0.41$\arcsec$ pix$^{-1}$ at ESO and 0.69$\arcsec$ pix$^{-1}$ at OHP. Exposure times of
1.1, 1.6, 1.1 and 0.8 hours were used to observe Arp 314, VV 304, LGG
455ab and LGG 455c, respectively; for the rest of the sample, the
total exposure time was 2 hours. In Table \ref{observinglog} we summarize the instrumental setup used in the observations.

\subsection{Data Reduction}
\label{datareduction}

The Fabry-Perot data were reduced by using the package developed by
Daigle et al. (2006). One of the main advantages of this reduction
package consists in the use of an adaptive spatial binning,
based on the 2D Voronoi tessellation method, applied to the 3D data
cubes. Details about data reduction using this package can be found in
Epinat et al (2008).

For the data cubes used in this study, an adaptive spatial binning was
applied to the data, in order to recover information on regions having
low signal-to-noise ratios (SNR). This allows optimizing the
spatial resolution to the SNR of the data. The adaptive binning allows
reaching an uniform SNR over the whole field-of-view with the highest
possible spatial resolution.  Indeed, with the spatial adaptive binning
technique, a bin is aggregating new pixels until it has reached a given
level that is set \emph{a priori}, called the signal-to-noise target
(SNRt). For each bin, the noise is determined from the r.m.s of the
continuum (the line-free region of the whole spectrum). The SNR is thus
the ratio between the flux in the line and the r.m.s. in the continuum.
Starting from an initial SNR, a pixel spectrum may be binned with
spectra of neighbour pixels to yield a new pixel of larger size (called
a bin) and of larger SNR. Disk regions of initial SNR higher than SNRt
(e.g. in the inner galaxy regions, spiral arms, star forming regions)
are not binned, maintaining the angular resolution as high as possible.
On the other hand, the angular resolution in disk regions with initial low
SNR (e.g. disk outskirts, interarm regions) is decreased during the binning
process in order to obtain an increase in SNR. Results presented here have been obtained with SNRt=5 per bin
for HCG compact group. In the case of the compact groups VV 304, LGG 455 and Arp 314 (observed at ESO), we used a SNRt=6, in order to avoid some artefacts present in the data at low SNR.

In order to quantify the mean SNR of a given data cube and its associated set of 2D maps, we have defined the index called SNRi (for signal-to-noise ratio index).  It measures,  above a certain flux threshold defining the galaxy area, the average number of pixels that need to be binned in order to reach the desired total SNRt. SNRi is thus defined as the total number of pixels divided by the total number of bins. As an example, if the image contains 10,000 pixels and 1,000 bins, SNRi=10 and, on average, 10 pixels are needed to be aggregated to form a bin for
a given SNRt. The smaller the value of SNRi, the higher is the mean SNR of that region; SNRi could not be lower than unity given that SNRi=1 means that an individual pixel has a SNR equal or higher than SNRt and no binning is necessary (and in that case the number of bins is equal to the number of pixels).

The OH sky lines were extracted by creating a data cube of the regions where no galaxies were
located. Wavelength calibrations were obtained by scanning the Ne
6598.95 $\rm{\AA}$ line under the same conditions as the observations.
In the end of the reduction process, we obtained the velocity field, the H$\alpha$ monochromatic,
the continuum and the velocity dispersion map of each galaxy. 

Velocity dispersion maps have been corrected from instrumental broadening. In the case of OHP galaxies (see section \ref{observations}), we have used the instrumental dispersion map derived from the Neon calibration lamp. For ESO objects (see section \ref{observations}), due to a lower SNR in the calibration data, we derived a mean instrumental correction over the whole field (which corresponds to $\sigma_{inst}$=5.6 km s$^{-1}$). We note that in the latter case the instrumental broadening does not change more than 10\% over the extension of the galaxies.  Under the assumption that the observed and instrumental profiles can be fitted with Gaussian functions, the actual velocity dispersion $\sigma$ can be
obtained using $\sigma=\sqrt{\sigma_{observed}^2-\sigma_{inst}^2}$. For a comparison, in Table \ref{kinematicinfo} (column 8) we list the mean velocity dispersion for each galaxy (values that are corrected for $\sigma_{inst}$). In the case of the groups HCG 14, 57 and 69, which were observed
twice, we have added the data, in order to
increase the SNR and the different maps were derived
from the added H$\alpha$ data cubes. Given the low-level emission
detected for HCG 44a and HCG 61, we have masked the velocity maps by
using the monochromatic images.

\subsection{Flux calibration}
\label{fluxcalibration}

\begin{figure}
\includegraphics[width=\columnwidth]{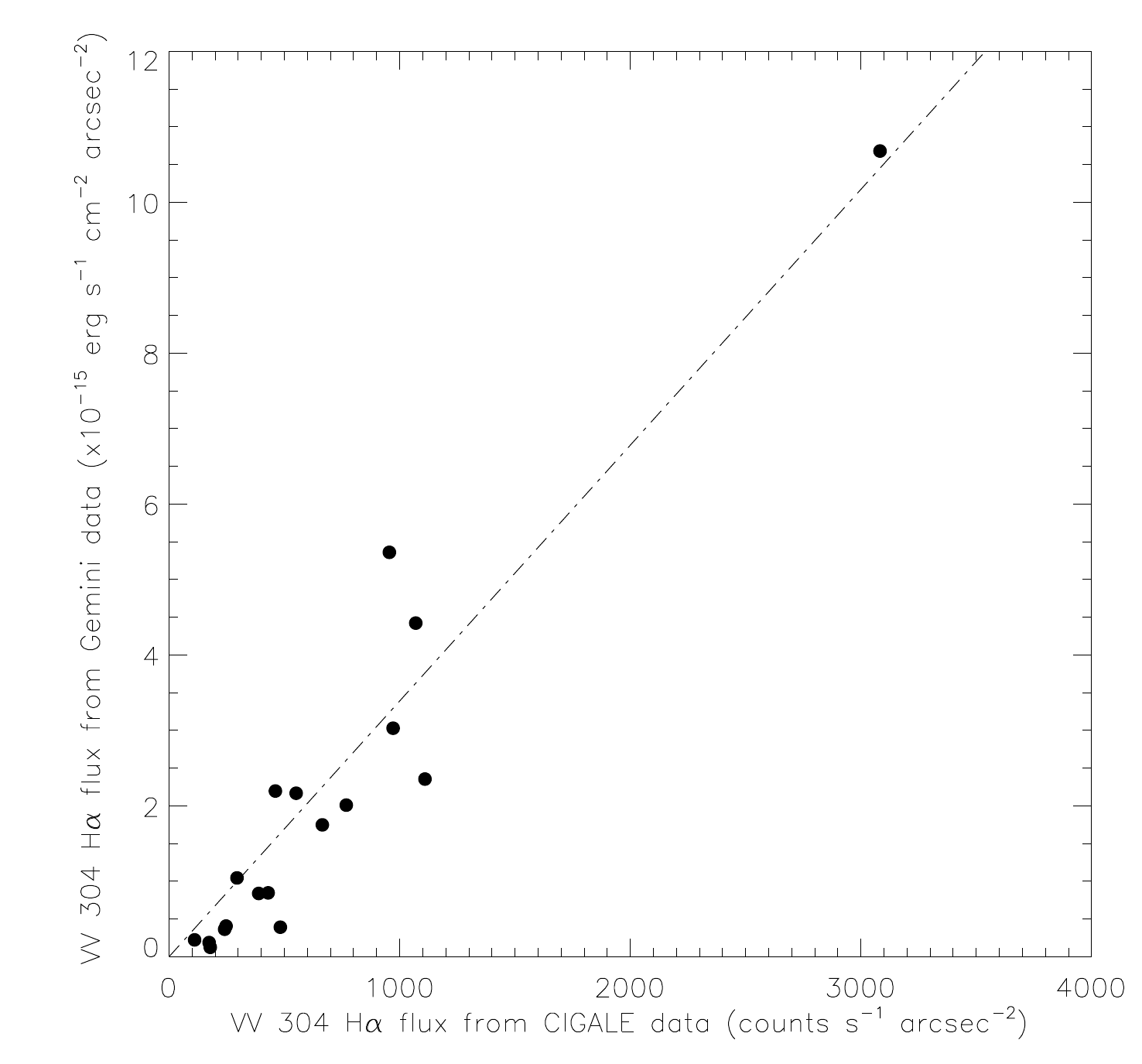}
\caption{Comparison between the H$\alpha$ fluxes measured from the monochromatic Fabry-Perot maps versus the H$\alpha$ fluxes derived from the Gemini/GMOS multislit observations of VV 304. The dashed-dotted line shows a linear regression on the data, where the zero point has been fixed to zero.}
\label{flux}
\end{figure}

The Fabry-Perot data described above have been obtained mainly to study the kinematic of compact groups galaxies. For this reason, we have taken no data for flux calibrators. However, an indirect calibration of the Fabry-Perot data, taken at taken at ESO, has been made using new Gemini GMOS multislit observations of the system VV 304. These data were observed during the program GS-2013B-Q-27 (PI: S. Torres-Flores), with a resolution that enabled us to separate the [NII] lines from the H$\alpha$ emission. Data reduction was performed using the Gemini Data reduction package in {\sc iraf}. In these observations the slit width was set to 1 arc sec, and the length of the slit was changed depending on the size of each source. In order to do the flux calibration, we have used a sample of 18 sources detected in the spiral arms of VV 304, which displayed typical spectra of a HII regions. For each source we have measured the H$\alpha$ flux by using the task {\sc splot} in {\sc iraf}. In order to compare these H$\alpha$ fluxes (erg s$^{-1}$ cm$^{-2}$) with the H$\alpha$ emission coming from the Fabry-Perot data (counts per seconds), we have used the monochromatic map produced in the data reduction process. In that map, we measured the H$\alpha$ emission in the same extraction windows that were defined in the multislit data reduction. The result of this analysis is shown in Figure \ref{flux}, where we adjusted a linear fit to the data. In this case we have set the zero point equal to zero. The resulting fit gave us a coefficient (slope) of 3.39$\times$10$^{-18}$ erg s$^{-1}$ cm$^{-2}$ counts$^{-1}$ s. Finally, this value was used to calibrate the Fabry-Perot monochromatic maps of VV 304, LGG 455 and Arp 314, assuming that LGG 455 and Arp 314 were observed under the same conditions as VV 304 (which was the case - the nights were photometric in all cases, and they were taken in subsequent nights, with the same instrumentation. In the case of compact groups observed at OHP we have attempted to do a flux calibration using the H$\alpha$ maps published by Vilchez et al. (1998), however these images do not provide a lower limit for the shown flux. Therefore, it was not possible to use these data as calibrators.

\subsection{Kinematic parameters and rotation curves}

For each galaxy, the kinematic position angle, inclination and systemic
velocity and rotation curve were estimated by fitting a modeled velocity
field to the observed data and minimizing the residual velocity
dispersion, following the method described in Epinat et al. (2008).
Due to the fact that the velocity fields of these perturbed systems are
complex, the centre of the models were not left free but fixed by using
the morphological centre of the galaxies defined as the peak of the
continuum image (as done by Torres-Flores et al. 2010). This procedure allowed us
to reduce the number of free parameters in the computation of the
rotation curves of the galaxies. We note that some groups have been observed under poor seeing conditions. This fact is not a problem in the case of the velocity fields that we derived, given that each bin corresponds to a given signal-to-noise ratio independently of the size of the seeing disk. Also, the rotation curves were plotted taking into account the seeing value (see Epinat et al. 2008).

For 27 galaxies out of the 42 in our sample, it was not possible to derive a
rotation curve, mainly for two reasons: (1) a few galaxies do not show
enough H$\alpha$ emission, (2) a few galaxies have a velocity field that is too
perturbed; in both cases it is impossible to obtain the rotation
curve based on the warm ionized gas component. Due to the fact that they are
not dominated by circular motions, perturbed velocity maps can not be
modeled in a reliable way, they produce unrealistic estimations for the
kinematic inclinations and position angles of the major axis of the
galaxy. On the other hand, less perturbed velocity fields need
nevertheless specific treatment. This is the case for velocity fields
for which the kinematic position angle is affected by a strong bar that
dominates most of the emission within the optical disk (e.g. HCG 44d).
In such cases, it is necessary to fix the position angle of the major
axis by using the morphological value determined outside of the bar. In
other cases the computation of the kinematic inclination yielded a value
close to that for an unrealistic face-on galaxy (e.g. HCG 1ab, HCG 53c and VV 304b). In those cases, we
have to fix the inclination of the galaxy by using its morphological
value (e.g. i$_{morph}=44$ degrees for HCG 1ab, i$_{morph}=52$ degrees for HCG 53c and i$_{morph}=46$ degrees for VV 304b). We also
had to fix the inclination by using the morphological value for
galaxies having a too irregular morphology which makes it difficult to get
a kinematic measurement (e.g. NGC 7232B). In Table \ref{kinematicinfo} we list
the main kinematic parameters for the galaxies for which we measure H$\alpha$ emission,
together with their respective morphological parameters (i.e. position angles and inclinations). In the same Table we include late-type spiral galaxies for which we do not detect H$\alpha$ emission. We note that in several cases it was not possible to derive the kinematic parameters (as discussed above), however, we have included these objects in Table \ref{kinematicinfo} in order to show their SNRi estimations.

In order to quantify the disagreement between both sides of the rotation curves, we have estimated the mean velocity difference for each curve. In most of the cases both sides of the rotation curves do not reach the same radius. In addition, the sampling of both sides is not the same (given the variations in the H$\alpha$ emission across of the galaxy). For these reasons, we have quantified the disagreement between both sides only up to the last radius for the less-extended side of the rotation curve, where we have matched the spatial sampling of both sides. In Table \ref{kinematicinfo} we list the asymmetries in the rotation curve for each galaxy. In this table we have also included the maximum rotational velocity for each galaxy, which was obtained from the rotation curves. This rotational velocity was used to obtain the dynamical mass for each galaxy, which was estimated by assuming that the mass has a spherical distribution. We have computed the dynamical mass of each galaxy at their optical radius and also at the last observed point in the rotation curve  (in the more-extended side of the curve). These values are listed in Table \ref{tablemasses}.

\begin{table*}
\begin{minipage}[h!]{\textwidth}
\caption{Kinematic properties of the sample. In this table we have included all the galaxies for which we expect H$\alpha$ emission following their morphological types.}
\label{kinematicinfo}
\centering
\renewcommand{\footnoterule}{}  % to avoid a line before footnotes
\begin{tabular}{lccccccccc}
\hline \hline
(1)\footnotetext{
(1) Galaxy ID;
(2) Position angle deduced from the
velocity field derived in this work;
(3) Optical position angle taken
from HyperLeda. The position angle is the position angle of the major
axis of the 25 mag arcsec$^{-2}$ isophote, in the B-band. In the case of
HCG 57d, no position angle estimation was found and we estimated it by
visual inspection of the optical image;
(4) Inclination deduced from the analysis of the velocity field derived in this work. An asterisk
indicates a case when we fixed the inclination by using the morphological value (column 5);
(5) Morphological inclination taken from HyperLeda. In the case of the HCG galaxies, this value was taken from
the axial ratio (cos(b/a)=i);
(6) Systemic velocity deduced from the velocity field analysis;
(7) Maximum rotational velocity derived from the rotation curve;
(8) Disagreement (asymmetry) between both sides of the rotation curve;
(9) Mean velocity dispersion ($\sigma$) in the whole field of the galaxy, corrected for instrumental width. In cases where the level of H$\alpha$ emission is low (as confirmed by the SNRi), we list the $\sigma$ value as \textit{nd} (no detection). In the case of HCG 1a and 1b, we list the mean velocity dispersion of the H$\alpha$ emission detected in the bridge between them: 
(10) SNRi index. It measures the averaged number of pixels per bin thus indicates the mean signal-to-noise ratio in the 2D maps (see section \ref{datareduction}): lower is this value, higher is the mean SNR, it is quality index. In cases where we do not detect H$\alpha$ emission, we list the SNRi value as \textit{nd} (no detection).
}
&(2)&(3)&(4)&(5)&(6)&(7)&(8) &(9)&(10)\\
Galaxy & PA$_{\rm{kin}}$ & PA$_{\rm{morph}}$ &i$_{\rm{kin}}$ &i$_{\rm{morph}}$ & V$_{\rm{sys FP}}$ & V$_{\rm{max}}$ & Asymmetry  & Mean $\sigma$ & SNRi\\
& deg & deg & deg & deg &km s$^{-1}$ & km s$^{-1}$ & km s$^{-1}$ &km s$^{-1}$ &\\
\hline
HCG 1ab              &   43    &  50   &    10*   &  44  & 10001$\pm$6 & 344$\pm$76 & 28 &25  &19.1\\
HCG 14a              &    ...    &     &    ...    &   &... &  ... & ... & \textit{nd}  & \textit{nd}\\
HCG 14c              &    ...    &     &    ...    &   &... &  ... & ... & \textit{nd} & \textit{nd}\\
HCG 25a              &    146$\pm$2     &    147 &    65$\pm$3    & 63  & 6231$\pm$2  &  170$\pm$9 & 23 &26 & 4.2\\
HCG 44a              &    ...    &     &    ...    &   &... &  ... & ... & \textit{nd} & 20.1\\
HCG 44c              &    137$\pm$1     &    131 &    53$\pm$3    & 54  &1228$\pm$2 &  163$\pm$11 & 8 & 20 & 16.6\\
HCG 44d              &    ...     &     &    ...    &   &... &  ... & ... & 25  & 8.5\\
HCG 53a              &     92$\pm$1     &    93  &    72$\pm$1    & 70 & 6206$\pm$1   & 266$\pm$8 &  13 & 24 &11.1\\
HCG 53c              &    104$\pm$4     &    122 &    52*    & 52  &6183$\pm$2   &  69$\pm$21 & 6  &26  &7.3 \\
HCG 57a              &    ...    &     &    ...    &   &... &  ... & ... &\textit{nd}  & \textit{nd}\\
HCG 57b              &    ...    &     &    ...    &   &... &  ... & ... & \textit{nd}  & \textit{nd}\\
HCG 57d              &     28$\pm$2     &    25  &    61$\pm$8    & 26 &8857$\pm$3   &  121$\pm$12 &  9 &26  & 4.2\\
HCG 61c               &    ...     &     &    ...    &   &... &  ... & ... &\textit{nd} &20.1\\
HCG 69a              &    ...     &     &    ...    &   &... &  ... & ... &24   &15.4\\
HCG 69b              &    ...     &     &    ...    &   &... &  ... & ... & \textit{nd}  &\textit{nd}\\
HCG 93b              &     155$\pm$4   &   39  &    29$\pm$18    & 70  & 4676$\pm$7 &  364$\pm$203 & 45 &23 & 10.4\\
HCG 93c              &    ...   &     &    ...    &   & ... &  ... & ... & \textit{nd} & \textit{nd}\\
VV 304a              &     104$\pm$2  &      120    & 39$\pm$12      & 42      & 3811$\pm$3  &   245$\pm$66 & 20 & 27 &4.9\\
VV 304b              &     58$\pm$5   &       33    & 46*   & 46      & 3867$\pm$7   &  207$\pm$79 &  70 & 25 &6.9\\
NGC 7232             &    108$\pm$2  &   100    & 64$\pm$7    &  86 & 1810$\pm$3  &  194$\pm$14 &  27 & 26  &12.5\\
NGC 7232B        &     162$\pm$8  &   1   & 32*   &  32      & 1775$\pm$2  &  68$\pm$48 & 27 & 28  &20.1\\
NGC 7233           &    14$\pm$2     &    ...   & 85$\pm$4    &  29 &1848$\pm$8   &  85$\pm$8 & 8 & 51 &1.0\\
Arp 314-1            &  115$\pm$2  &   33    &  18$\pm$14    & 48     & 4042$\pm$1 & 159$\pm$116  & 26 & 27  &4.9\\
Arp 314-2            &     70$\pm$6 &   ...    &   75$\pm$23 &  22        & 4031$\pm$7  & 210$\pm$24 & 32 & 30  &4.9\\
Arp 314-3            &   130$\pm$4  &  175   &   70$\pm$18   & 90        & 3757$\pm$2  &  94$\pm$12 & 10 & 25  &15.0\\
\hline
\end{tabular}
\end{minipage}
\end{table*}

\begin{table*}
\begin{minipage}[h!]{\textwidth}
\caption{Dynamical masses for the galaxies of this sample}
\label{tablemasses}
\centering
\renewcommand{\footnoterule}{}  % to avoid a line before footnotes
\begin{tabular}{lccccc}
\hline \hline
(1)\footnotetext{
(1) Galaxy ID;
(2) Maximum radius reached by the rotation curve;
(3) Optical radius of the galaxy;
(4) Maximum rotational velocity for each galaxy;
(5) Dynamical mass estimated at R$_{max}$;
(6) Dynamical mass estimated at R$_{25}$;
}
&(2)&(3)&(4)&(5)&(6)\\
Galaxy & R$_{max}$ & R$_{25}$ & V$_{\rm{max}}$  & Mass at R$_{max}$ & Mass at R$_{25}$ \\
 & kpc & kpc & km s$^{-1}$ & $\times$10$^{10}$ M$_{\odot}$ & $\times$10$^{10}$ M$_{\odot}$ \\
\hline
HCG 25a            &    18.13     &  15.81  &   170$\pm$9     &  12.0 &   11.0   \\
HCG 44c            &     5.82     &  6.786  &   163$\pm$11    &  3.6 &  4.2   \\
HCG 53a            &    32.93     &  30.81  &  266$\pm$8      &  54.0 &  51.0   \\
HCG 53c            &     5.87     &  9.438  &   69$\pm$21     &  0.6 &  1.0   \\
HCG 57d            &     14.5     &  7.557  &   121$\pm$12    &  4.9 &  2.6   \\
HCG 93b            &    22.28     &  26.39  &   364$\pm$203   &  68.0 &  81.0   \\
VV 304a            &    16.75     &  18.06  &    245$\pm$66   &  23.0 &  25.0  \\
VV 304b            &     17.6     &  18.87  &   207$\pm$79    &   18.0 &  19.0   \\
NGC 7232           &     8.02     &  10.32  &   194$\pm$14    &  7.0 &  9.0   \\
NGC 7232B          &     6.34     &   7.64  &   68$\pm$48     &  0.7 &  0.8   \\
NGC 7233           &     1.81     &  6.503  &   85$\pm$8      &  0.3 &   1.1   \\
Arp 314-1          &     6.58     &  8.175  &  159$\pm$116    &  3.9 &  4.8  \\
Arp 314-2          &    10.73     &  9.775  &  210$\pm$24     &  11.0 &      10.0  \\
Arp 314-3          &     11.7     &  8.207  &   94$\pm$12     &  2.4 &  1.7   \\
\hline
\end{tabular}
\end{minipage}
\end{table*}

\section{Kinematic description of each interacting system}

In this section we describe the main kinematic features for the
members of the studied groups. In Appendix A (online data only, Figs. A1 to A14) we show, for each observed
field, an optical DSS image (top left), a velocity field (top right),
an H$\alpha$ monochromatic image (bottom left) and a dispersion map
(bottom right). In Appendix B (online data only, Fig. B1) we
show a rotation curve for each galaxy, when it was possible to
derive it. As an example, in Fig. \ref{vfvv304_ned} (included in the electronic version of this paper) we shown the different maps that we derived for the members of the system VV 304, and in Figure \ref{rc1} (also included in the electronic version) we shown the rotation curves that we derived for galaxies VV 304a and VV 304b. The distances to the groups were taken from NED. Distances were corrected for the Virgo, Great Attractor and Shapley supercluster infall (Mould et al. 2000).

\subsection{HCG 1}

Fig. A1 shows the various maps obtained
for HCG 1a and 1b. The H$\alpha$ map of HCG 1 reveals  a diffuse
H$\alpha$ emission for these two galaxies, although they
are respectively of Sc and Im type. The galaxies
HCG 1c and 1d do not have any H$\alpha$ emission, which is
consistent with their morphological types (E0 and S0, respectively). One should note that the transmission of the interference filter we used was quite low ($\sim$40\%). Interestingly, the H$\alpha$ emission shown in Fig. A1
is mainly concentrated in the bridge connecting HCG 1a and HCG 1b, 
also detected in the optical images of this system (see Hickson 1993). Since the 3D shape of the group is not known, it is 
not possible to tell if material is flowing along this bridge. Despite the irregular H$\alpha$ morphology of this system, a velocity field was derived. The rotation curve derived from this map (see Fig. B1) shows a disagreement in the inner part, however, in the outer part both sides match. In this case, we measure an asymmetry of 28 km s$^{-1}$.

\subsection{HCG 14}

No H$\alpha$ emission was detected in this group, despite a long 
exposure time and the relatively late morphological types of two of its galaxies. However, as for HCG 1cd, we note that the transmission of the interference filter we used was quite low (in this case $\sim$25\%). 

\subsection{HCG 25}

We observed four members of the group (HCG 25a, b, f and g) 
in the same field of view but detected H$\alpha$ emission only for HCG 25a.

The H$\alpha$ emission map (Fig. A2) suggests that
the northern side of the galaxy is forming stars more actively than
the southern side, with a prominent northern spiral arm. However 
the brightest star forming region is located on the southern side.
The velocity field shown in Fig. A2 is fairly regular,
in agreement with the general morphology of this galaxy.
The velocity dispersion map shows a mean value of $\sigma$=26 km s$^{-1}$.

The rotation curve of HCG 25a, shown in Fig. B1, 
reaches values of $\sim$180 km s$^{-1}$. Despite the regular 
grand design of its velocity field, the rotation curve of this galaxy
shows some differences between the approaching and receding sides. 
On average, this disagreement reaches a value of 23 km s$^{-1}$.  
Nevertheless, the rotation curve increases for both sides 
out to the optical radius of the galaxy, where it reaches a maximum. 
A small offset of the kinematical centre (2 arcsec, i.e. 0.8 kpc, to the North-West)  
could reconcile the two sides of the rotation curve in their rising part,
then resulting in a velocity difference of about 30 km s$^{-1}$ between
the two sides of the curve in the outer parts. 
Considering the high inclination of the galaxy (65 degrees) this 
apparent offset could be due to dust extinction.

\subsection{HCG 44}

Three different fields have been observed for HCG 44, 
covering members HCG 44a, c and d respectively. The galaxy HCG 44a 
shows little H$\alpha$ emission in the centre and in two blobs
(at a very low level) as can be seen in Fig. A3.
No velocity field can be obtained from our data. HCG 44c shows
H$\alpha$ emission distributed along a ring (Fig. A4) 
with a strong peak at the centre of the galaxy (as expected for a
Seyfert 2 galaxy) and a brighter emission on both ends of the
major axis. This H$\alpha$ annular emission coincides with 
the UV \textit{GALEX} emission of this galaxy (see Gil de Paz et al. 2007). 
The ring seen in our H$\alpha$ image could be the result of the 
overlaping of two tight spiral arms of this galaxy. 
Its velocity field is quite regular, with no signatures of any
interaction. The position angle of the kinematical major axis is well
defined, together with the isovelocity lines across its main body. 
The rotation curve of this object reflects its ring-like structure,
as can be seen in Fig. B1. In the inner first kpc, the rotation
curve increases strongly, almost linearly. Between 1 and 2 kpc there 
is no information, given the lack of H$\alpha$ emission there. From 2 to 4
kpc the curve keeps increasing gently, reaching a maximum value of about
160 km s$^{-1}$, close to the optical radius, with a very good agreement 
between approaching and receding sides of the galaxy, which produces a mean asymmetry of 8 km s$^{-1}$ between both sides.

The H$\alpha$ emission map of HCG 44d shows several strong emitting
knots along a bar-like structure. Strangely, this galaxy was not classified 
as barred by Hickson (1993) although the B and R-band images 
(from the Hickson's catalogue and available in the NED database)
show a prominent bar. The two arms starting at the end of the bar 
are exceedingly open, suggesting that their origin could be linked 
to a past or ongoing tidal interaction.  Two emitting sources
are conspicuous in the northeastern spiral arm whereas only diffuse 
H$\alpha$ emission can be seen in the southwestern arm (Fig. A5). The velocity field of HCG 44d is dominated by
the bar. The isovelocities are almost parallel to the bar. The velocity amplitude along 
the kinematical major axis reaches a value of $\sim$160
km s$^{-1}$. The velocity dispersion map reaches its highest values all over
the bar, ranging from $\sigma_{observed}\sim$25 to 40 km s$^{-1}$. 
Because of the strong bar and the tidal arms, it was not possible to 
draw a reliable rotation curve for this galaxy, although we fixed the 
position angle by using its morphological value.

\subsection{HCG 53}

Our Fabry-Perot data cover galaxies HCG 53a, 53b and 53c but H$\alpha$ emission
was found only in a and c.

The monochromatic emission distribution in HCG 53a reaches the optical radius 
of the galaxy. It is asymmetric (see Fig. A6) 
with bright H$\alpha$ emission regions on the eastern side. 
Interestingly, the \textit{GALEX} data of this galaxy show an 
homogeneous emission across the whole galaxy. Note also that 
 HCG 53 was not detected in the infrared by Allam et al. (1996).
The velocity field has been derived with good quality across the whole galaxy 
(see Fig. A6). It is regular on the eastern side and 
slightly warped towards the north, on the western side.
The velocity amplitude is roughly $\pm$260 km s$^{-1}$.  
There is an excellent agreement between the optical and the kinematical
position angles. The rotation curve of HCG 53a is quite regular 
(see Fig. B1) with a good agreement between the approaching 
and receding sides and it is almost flat, although very slightly increasing beyond 5 kpc.

The small galaxy HCG 53c has a strong bar, which is clearly seen on the DSS
image and on the R-band image of the Hickson's catalogue. Along this structure we detected several bright knots containing almost half of the total H$\alpha$ 
emission of the galaxy (see Fig. A6). 
The brightest H$\alpha$ emission comes from the south end of the bar. 
However, no H$\alpha$ knot is observed on the north end of the bar, 
which could mean that the star-formation is not linked to a resonance. 
Another H$\alpha$ emitting knot coincides with the centre of the galaxy. 
The bar and the kinematic minor axis position angles have an angular 
separation of 48 degrees (157 and 25 degrees, respectively) leading to 
a velocity gradient along the bar of $\sim$50 km s$^{-1}$ and the
strong S shape of the bar signature can be seen in the velocity field. 
The velocity amplitude along the major axis of the galaxy is $\sim$90 km s$^{-1}$.
HCG 53c has an increasing rotation curve (Fig. B1) 
reaching $\sim$70 km s$^{-1}$ at 6 kpc radius. 
For this object, we adopted the inclination value suggested by its morphology (52 degrees).

The velocity dispersion is quite homogeneous across the whole disk of both
HCG 53a and 53c, with values of $\sigma$=24 and 26 km s$^{-1}$ respectively.

\subsection{HCG 57}

Galaxies HCG 57a, b, c, d, e, g and h were observed in the same field of
view (member f was about 1' outside the field). Two observations were made,
on two successive nights, with different interference filters. 
The first observation, with rather bad seeing, was made through a filter 
centred at 673.5 nm (FWHM 1.0 nm) best suited for components 57b, c, d and e.
The second observation, with good seeing, was made through a filter centred 
at 672 nm (FWHM 1.0nm) better suited for HCG 57a but does not show any 
clear H$\alpha$ emission despite these good conditions.
Only member d displays strong H$\alpha$ emission all over its disk 
(see Fig. A7) even through the worse-suited filter was used. 
No H$\alpha$ emission could be seen for HCG 57c, e, g and h and there is 
but a suspicion of faint emission on both sides of HCG 57b (northwest and southeast), 
possibly coming from the spiral arms. 
Also \textit{GALEX} data clearly show UV emission for member d only 
and some diffuse \textit{GALEX} UV emission can be seen for member b.

HCG 57 has been observed in mid-IR by Bitsakis et al. (2010) showing that
galaxy d has, by far, the highest SFR of the group, which is consistent 
with our H$\alpha$ observation.

The Spitzer IR spectrum of HCG 57a suggests a non-star-forming mechanism able 
to excite the H2 in the disk of the galaxy, showing that this group may be in 
a specific phase of rapid transformation (Cluver et al. 2013). 
HCG 57a is classified as an AGN by Martinez et al. (2010) 
and as a LINER (low-ionization nuclear emission-line region galaxies) by Gavazzi et al. (2011). 
HCG 57d is classified as a low-luminosity AGN that coexists with circumnuclear
star formation by Martinez et al. (2010).

The H$\alpha$ emission is strong along the two main spiral arms of HCG 57d,
the brightest being the southern arm, which is in the direction of 
HCG 57a. Almost no H$\alpha$ emission can be seen in the centre of this galaxy. 
A diffuse H$\alpha$ structure extends along the southern spiral arm, connecting
HCG 57d with the central region of HCG 57a. We note that the spectrum of 57c and 57g analysed by Martinez et al. (2010) do not show the presence of the H$\alpha$ emission line, which is consistent with our results. On the other hand, Martinez et al. (2010) found H$\alpha$ emission for members HCG 57b, 57e and 57h, for which we have detected no emission. This fact may be linked with the detection limit of our current observations.

The velocity field of HCG 57d is fairly regular, with an amplitude of 
$\sim$200 km s$^{-1}$ and no clear signature of interaction can be seen
despite the apparent proximity of the distorted galaxy HCG 57a. 
However, there is a small change in the position angle of the major axis 
along the radius (see Fig. A7). Also, the velocity field
is influenced by the H$\alpha$ extension that seems to connect HCG 57d with 
the central region of HCG 57a. Finally, the rotation curve of HCG 57d shows a 
good agreement between the approaching and receding sides up to 6kpc from 
the centre but both sides do not match in the outer parts. It reaches a 
maximum velocity of $\sim$120 km s$^{-1}$ and can be traced beyond
the optical radius, where it becomes flat.

The highest values of velocity dispersion are reached along the spiral arms
of HCG 57d, following the same ring-like structure (see Fig. A7). 
The mean value for the whole velocity dispersion map is $\sigma$=26 km s$^{-1}$.

\subsection{HCG 61}

This group, also known as ``The Box", is a triplet formed by HCG 61a, c and d.
Allam et al. (1996) confirm the presence of dust and infrared emission in HCG61c.
Some H$\alpha$ emission can be seen in the southeastern side of HCG61 c, corresponding 
to the brightest region in the B and R-band optical images (see Fig. A8). 
HCG 61a and d are S0 galaxies for which no H$\alpha$ emission is expected. However,
the central regions of both galaxies display some weak H$\alpha$ emission, in agreement 
with the H$\alpha$ maps from Vilchez et al. (1998).
Due to the low SNR of the H$\alpha$ emission, no rotation curve could be computed for any member of
this group.

\subsection{HCG 69}

HCG 69 is a quartet composed of two late-type galaxies and two lenticulars. 
HCG 69a shows a strong dust lane across its major axis.

For this group, we have observed twice the same field of view, covering the four
members of this system and the two data cubes were added. Very faint H$\alpha$
emission can be seen for member a, fainter still along the dust lane, as expected (see Fig. A9), 
whereas no emission at all was found for members b, c and d. 
\textit{GALEX} ultraviolet emission is bluer for HCG 69a than for HCG 69b and no UV
emission is detected for HCG 69c and 69d, as expected from their types.
Because of the low SNR of the H$\alpha$ emission, no rotation curve could be computed for any galaxy of
this group.

\subsection{HCG 93}

Two fields were observed for this group, one for HCG 93a and c and the other one
for HCG 93b. No H$\alpha$ emission has been found for HCG 93a and 93c, as expected 
from their morphological types. We note that the transmission of the interference filter that we have used was low, close to a $\sim$25\%.
Mendes de Oliveira et al. (1994) mention a stellar extension of HCG 93c towards 93a, 
suggesting a past interaction.

Despite the low transmission of the interference filter, we detected H$\alpha$ emission in the late-type galaxy HCG 93b 
(consistent with Torres-Flores et al. 2010), with bright regions on the tips of the bar 
and on the inner region of the western spiral arm (see Fig. A10). However, the velocity field of HCG 93b 
is so perturbed, with a strong change of the position angle along the major axis, 
that we could not compute any rotation curve, even when fixing the kinematical parameters 
by using their morphological parameters. The stretching of the spiral arms and the 
perturbation of the velocity field, which cannot be explained by the presence of a 
strong bar alone, suggest that HCG 93b is interacting with the other galaxies of the group.

\subsection{VV 304}

VV 304 is a close pair of galaxies inside a quartet (Vorontsov-Velyaminov 1959). 
VV 304a and VV 304b exhibit perturbed spiral arms on the DSS optical image.
For both galaxies, we detect diffuse H$\alpha$ emission in a ring-like structure 
on which bright HII regions are superimposed (see Fig. \ref{vfvv304_ned}; Fig. A11 in Appendix A). 
Almost no H$\alpha$ emission is detected in the nuclei of these galaxies 
where the optical DSS images display strong emission. Both monochromatic images 
are much alike (see Fig. \ref{vfvv304_ned}) but VV 304b is a bit brighter. 
One can see some bright H$\alpha$ knots in the centre of the pair, where the disks
seem to overlap, which could be the result of the interaction.

Despite the interaction, the velocity field of VV 304a looks fairly regular. 
However, some asymmetries in the isovelocity lines can be seen along the minor axis. 
Also, this galaxy has a peculiar rotation curve (see Fig. \ref{rc1}). 
Although the agreement is satisfying between the approaching and receding sides, 
several bumps can be seen on both sides of the rotation curve. Despite these bumps, 
the curve remains fairly flat out to 15 kpc, where no more H$\alpha$ emission is detected 
(note that the approaching side is more extended because of the overlapping of the disk 
of VV 304b on the receding side of VV 304a). The velocity dispersion map of VV 304a is rather smooth, 
without any peak, and displays a mean value of $\sigma$= 27 km s$^{-1}$.

The velocity field of VV 304b is clearly more perturbed than that of VV 304a, 
with a difference of almost 30 degrees between the optical and kinematical position angle of 
the major axis for this galaxy. The southwestern part of the disk (in the direction of VV 304a) 
displays almost constant radial velocities (about 300 km s$^{-1}$ lower than the velocities of 
the near side of VV 304a) whereas there is a regular velocity gradient along the major axis 
on the receding side. As a result, the rotation curve of VV 304b is completely asymmetric 
(see Fig. \ref{rc1}) confirming the interaction.
The velocity dispersion map for this object is similar to that of VV 304a.

\subsection{LGG 455}

Following Garcia's (1993) catalogue, this group is formed
by three spiral galaxies, NGC 7232, NGC 7232B and NGC 7233, and one
lenticular galaxy, IC 5181. Subsequently, Garcia (1995) catalogued this
group as a compact group.

Two different fields of view where observed for this group,
one for NGC 7232 and NGC 7233, the other one for NGC 7232B. The results
are given in Fig. A12 and Fig. A13 respectively.

The H$\alpha$ map of NGC 7232 shows faint diffuse emission all
over the disk, with two bright knots, one is close to the nucleus
(7.5$\arcsec$ west) and the second one is at 22$\arcsec$ west from 
the centre of NGC 7232. A faint tail-like structure can be seen 
eastward of NGC 7232. The velocity field of NGC 7232 shows a clear 
velocity gradient along the kinematical major axis, well aligned with 
the morphological one of the galaxy. The large bin sizes displayed in the velocity 
field of this galaxy are due to the low SNR of the H$\alpha$ emission. 
It results in a patchy appearance of the velocity field, however
no strong signature of interaction can be seen. The velocity dispersion map of this galaxy shows the highest values where
the monochromatic map peaks, at $\sim$50 km s$^{-1}$. Despite
the fairly regular velocity field of this galaxy, its rotation curve
is quite chaotic (see Fig. B1). In the inner part, 
there is no agreement between both sides of the rotation curve. 
The approaching side displays a large bump at 10$\arcsec$ before 
growing linearly, whereas the receding side exhibits a large velocity scatter
with increasing radius. Furthermore, the rotation curve hardly reaches
half the optical radius.

NGC 7233 shows a strong nuclear H$\alpha$ emission and a faint
diffuse H$\alpha$ emission across the whole optical disk, but no
H$\alpha$ knot on the disk, not even in the spiral arms visible
on the optical DSS images. 
The velocity field shown in Fig. A12 has been masked and limited 
to the very central part of the galaxy because of parasitic ghost images
in the outer parts (the velocity gradient was inverted at large radii, 
reminding of a problem encountered with HCG 2b by Torres-Flores et al. 2009). 
As a result, the rotation curve shown in Fig. B1 
is limited to the rising part.

Most of the H$\alpha$ emission of NGC 7232B can be seen 
in the bar and the departure of the southern spiral arm of this galaxy.
The velocity field of NGC 7232B displays a small velocity amplitude 
$\sim$80 km s$^{-1}$. 
The rotation curve of this galaxy (shown in Fig. B1) 
grows almost linearly with the radius. In the first 10$\arcsec$ 
the velocities display a large scatter, with poor agreement between
the approaching and receding sides. The latter climbs at a rotation
velocity of 70 km s$^{-1}$, almost reaching the optical radius.

\subsection{Arp 314}

Arp 314 is formed by three late-type spiral galaxies: 
Arp 314 NED01, NED02 and NED03 (hereafter Arp 314-1, Arp 314-2 and Arp 314-3). 
Garcia (1995) classified this triplet as a compact group (LGG 467). 
\textit{GALEX} UV images reveal a prominent tidal tail eastward from 
Arp314-2 and Arp314-3 (Gil de Paz et al. 2007). 
The \textit{GALEX} UV tidal tail can be seen on optical images 
such as the Palomar image given in NED or the \textit{SDSS} DR8 release, 
also showing an outer shell around Arp 314-1, which is another clear 
signature of interaction. Interestingly, this tidal tail points
toward a seemingly fourth small galaxy at about 4 arcmin to the east of Arp 314-3 
(however no redshift for this object can be found in the literature).

Our H$\alpha$ map of Arp 314-1 shows a complex shape (see Fig. A14). The most
conspicuous structure is a chain of bright HII regions in the centre,
apparently along a spiral arm. These regions exhibit large asymmetric 
H$\alpha$ profiles, with a mean width of $\sim$35 km s$^{-1}$  
($\sigma$). 
The brightest H$\alpha$ emitting region is located at the southwest
end of this chain, it is also prominent on the \textit{GALEX} UV images 
and optical \textit{SDSS} images. The H$\alpha$ profiles in this bright
region are fairly symmetric whereas those observed close to the nucleus
are more irregular, with a second component that can be seen on their 
redshifted side, but it is not detected across the whole disk of the galaxy. 
Despite the complex H$\alpha$ morphology of Arp 314-1, 
its velocity field is fairy regular, with a clear velocity gradient along
the kinematic major axis. However some regions show up with abnormal
velocities, on both sides of the galaxy and more especially on the 
southern edge of the disk. Also, we find a strong misalignment between the position angle of the
morphological major axis (33 degrees from the Hyperleda database) and 
the PA of the kinematical major axis (115 degrees from our veocity field). 
This discrepancy is a clear signature of interaction between galaxies 
(Torres-Flores et al. 2010). In Fig. B1 we show the
rotation curve of Arp 314-1. In the inner 10$\arcsec$ we find a strong
bump on the receding side, with rotation velocities climbing above 
250 km s$^{-1}$. Then both sides display a slower rotation velocity around
$\sim$150 km s$^{-1}$. Despite the clear signatures of interaction 
shown by Arp 314-1, its rotation curve remains fairly flat in the outer 
parts, with a good agreement between both sides.

Arp 314-2 also shows a complex H$\alpha$ morphology, with most of 
the H$\alpha$ emission in the nuclear region and no clear spiral arms.
The general pattern is alike that of the \textit{GALEX} UV images 
and optical \textit{SDSS} images of this galaxy.
The H$\alpha$ profiles observed in the central region are more symmetric 
than for Arp 314-1, with a $\sigma$ of $\sim$40 km s$^{-1}$. 
The other bright H$\alpha$ knots found in the disc of this object exhibit
symmetric H$\alpha$ profiles (with an average $\sigma$=30 km s$^{-1}$). 
Contrary to Arp 314-1 no secondary component is observed on any of the
profiles in Arp 314-2. The very central region of Arp 314-2 (inside the
first 7$\arcsec$ diameter) displays a normal velocity gradient, with no
signature of interaction. Outside these first few arc seconds, the
velocity field is much more perturbed and chaotic. On the western edge 
of the disk, one can see a large region with abnormally high velocity.
This feature causes a bump on the approaching side of the rotation curve 
(at $\sim$1 kpc), as shown in Fig. B1. 
This rotation curve shows a short extension for the receding side 
($\sim$4 kpc) whereas the approaching side reaches a radius of $\sim$12 kpc, 
beyond the optical radius. Both sides disagree strongly within 
the first ($\sim$4 kpc), with quite opposite behaviours. 
The outer part of the rotation curve is drawn by the approaching side
alone, displaying a solid-body shape.

Arp 314-3 looks like an irregular galaxy or alternatively it could 
be a tidal debris. A few low-intensity H$\alpha$ emitting
regions are detected in the body of this object. These regions have
well defined and symmetric H$\alpha$ profiles. The velocity field
shows an amplitude of about 120 km s$^{-1}$ and the resulting 
rotation curve (see Fig. B1) has a  
symmetric behaviour for both approaching and receding sides. 
It rapidly reaches an intermediate flat part with velocities around 
40 km s$^{-1}$ at the optical radius (4 kpc) then climbs almost linearly 
up to 120 km s$^{-1}$ at a radius of 12 kpc. The outer part of the rotation
curve must be taken with care however since it has been drawn assuming 
that Arp 314-3 is a rotating disk whereas the velocity field displays
isovelocity lines quite different from those expected with pure circular
motions. For instance, one can see unexpectedly high changes in radial 
velocities along the minor axis, suggesting streaming motions. Also, 
a strong warp of the disk could explain the apparently high rotation 
velocities reached by the outermost parts of the rotation curve.
Deep red or near infrared band images could help understanding
if Arp 314-3 is actually a star forming galaxy containing nevertheless 
an old stellar population or, alternatively, a recent tidal debris.

\begin{table*}
\begin{minipage}[t!]{\textwidth}
\caption{Interaction indicators based on H$\alpha$ emission. In this table we have included all the galaxies for which we expect H$\alpha$ emission following their morphological type. Description of the columns.
H$\alpha$ distribution:
(A1) Lack of H$\alpha$ emission which respect to what is expected taking into account the morphological type of the galaxy;
(A2) Complex H$\alpha$ structure;
(A3) Tidal tails;
(A4) Distorted spiral arms.
Velocity Field:
(B1) Highly disturbed velocity field;
(B2) Gaseous versus stellar major-axis misalignment;
(B3) Changing position angle along major axis.
Rotation Curve (RC):
(C1) Disagreement between both sides of the RC;
(C2) Unexpected solid body behavior for the RC;
(C3) Truncated rotation curve (shorter than R$_{25}$)}.
\label{indicators}
\centering
\renewcommand{\footnoterule}{}  % to avoid a line before footnotes
\begin{tabular}{lcccccccccc}
\hline
%\multicolumn{1}{c}{Galaxy} &  \multicolumn{10}{c}{Interaction indicator} \\
\hline
Galaxy&\multicolumn{4}{c}{H$\alpha$ distribution}&\multicolumn{3}{c}{Velocity field} & \multicolumn{3}{c}{Rotation curve}\\
&\multicolumn{4}{c}{---------------------------}&\multicolumn{3}{c}{--------------------}&\multicolumn{3}{c}{--------------------}\\
& A1&A2&A3&A4&B1&B2&B3&C1&C2&C3\\
\hline
HCG 1ab   & + & ... &  + & ... & + & -- & ... & + & + & + \\
%HCG 1b   & + & ... &  ... & ... & + & ... & ... & ... & ... & ... \\
HCG 14a   &+& ... & ...  & ... & ... & ... & ... & ... & ... & ... \\
HCG 14c   &+& ... & ...  & ... & ... & ... & ... & ... & ... & ... \\
HCG 25a   &--& + &  -- & -- & -- & -- & -- & + & -- & -- \\
HCG 44a   & + & ... & ...  & ... & ... & ... & ... & ... & ... & ... \\
HCG 44c   &-- & + &  -- & + & -- & -- & -- &  -- & -- & -- \\
HCG 44d   & -- & + &  -- & + & + & ... & ... & ... & ... & ... \\
HCG 53a   & -- & -- &  -- & -- & -- & -- & -- & -- & -- & --  \\
HCG 53c   & -- & -- &  -- & -- & -- & + & -- & -- & + & + \\
HCG 57a   & + & ... & ...  & ... & ... & ... & ... & ... & ... & ... \\
HCG 57b   & + & ... & ...  & ... & ... &...  & ... & ... & ... & ... \\
HCG 57d   & -- & + &  -- & + &  --  & + & + & + & -- & + \\
HCG 61c   & + & ... & ...  & ... & ... & ... & ... & ... & ... & ... \\
HCG 69a   & + & ... & ...  & ... & ... & ... & ... & ... & ... &...  \\
HCG 69b   &+ & ... &...   & ... & ... & ... & ... & ... & ... & ... \\
HCG 93b   & -- & -- & --  & + & + & + & -- & + &  -- & -- \\
HCG 93c   & + & ... & ...  & ... & ... & ... & ... & ... & ... & ... \\
VV 304a   &-- & + &  -- & + & -- & + & + & + & -- & +  \\
VV 304b   & --& + &  -- & + & + & + & + & + & -- & +  \\
NGC 7232  &--& + &  -- & ... & + & -- & -- & + & + & +  \\
NGC 7232B &--& + &  -- & + & + & + & ... & + & + & +  \\
NGC 7233  &--& + &  -- & + & + & ... & ... & -- & + & +  \\
Arp 314-1 &--& + &  + & + & + & + & + & + & -- & + \\
Arp 314-2 &--& + &  + & + & + & ... & + & + & + & --  \\
Arp 314-3 &--& + &  + & + & + & + & ... & -- & + & --  \\
\hline
\end{tabular}
\end{minipage}
\end{table*}

\section{Discussion and concluding remarks}

Kinematic information on interacting galaxies has been quite useful
in the determination of the evolutionary stages of compact groups.
Mendes de Oliveira et al. (1998), Amram et al. (2003), Plana et al.
(2003) and Torres-Flores et al. (2010) have listed a set of kinematic
interaction indicators, which may give insights to the interaction
history of galaxies, and hence to the evolutionary stages of the
groups to which they belong. For example, Amram et al. (2003)
suggested that highly disturbed velocity fields, double nuclei,
double kinematic gas components and high amplitude discrepancies
between both sides of the rotation curves imply strong galaxy-galaxy
interactions or mergers. On the other hand, stellar and gaseous
major axes misalignments and tidal tails suggest collisions that
may not always lead to merging. Here, we used these indicators to
determine the evolutionary stages of the different groups analysed
in this paper.  In Table \ref{indicators} we listed the different indicators for
each galaxy and in the following we discuss the results for each
group.

In {\bf HCG 1}, we detect H$\alpha$ emission in the bridge
between 1a and 1b and a diffuse emission in the centres of members a and b, which suggests that a strong interaction event has occurred
in this system. In Stephan's quintet, a group that is widely recognised
as being in an advanced stage of evolution, ionized gas is not
present in the centres of the member galaxies.  In this sense HCG
1 resembles the Stephan's quintet. We conclude that the star
formation in the bridge between HCG 1a and 1b may have been
strongly enhanced due to interactions.

The three observed galaxies in {\bf HCG 14} (members a, b and c)
have morphological types Sb, E5 and Sbc respectively. Despite two
of them being spiral galaxies, we do not detect any H$\alpha$
emission for the member galaxies of this group. The elliptical
galaxy HCG 14b was pointed out by Mendes de Oliveira and Hickson
(1994, their Fig. 4) to have a surface brightness profile which
is shallower than those of other elliptical galaxies of similar
luminosities, resembling a surface brightness profile of a cD galaxy (although HCG 14b
is an intermediate-luminosity galaxy). This shallow profile may
indicate previous merging events.  We suspect that this group could
be the result of one or more past mergers or accretion events, given
the shallow profile of HCG 14b, but this question remains open,
awaiting further observations

We observed three galaxies in {\bf HCG 25}, members a, b and f.
There are four other galaxies (c, d, e and g) originally catalogued
by Hickson (1982) to be members of the same group but which are
actually part of another system in the foreground.  Only for HCG
25a (an SBc galaxy) we detect H$\alpha$ emission.  We note that HCG
25b (an SBa) may be interacting with the small S0 galaxy HCG 25f,
given an optical bridge seen between these two objects.

We find a disagreement between both sides of the rotation curve of
HCG 25a, and this is usually taken as an indication that the galaxy
has suffered interaction. However, we find no other signature for
an ongoing merger or strong interaction event given that the velocity
field of this galaxy is almost undisturbed. We tend to favour an
early-stage-of-evolution scenario for this group although this
conclusion is drawn from the kinematical observation of only one
group member, HCG 25a, and from the inspection of the B and R images
of HCG 25bf. More data are necessary to allow a better definition
of the evolutionary stage of this triplet.

We observed three galaxies in group {\bf HCG 44}, members a, c and
d.  HCG 44a displays little H$\alpha$ emission in its centre, which
is expected given its morphological type, Sa.  No rotation curve
could be derived.  For HCG 44c, an SBc, we were able to derive the
rotation curve. Just one indicator is flagged positive in this case.
HCG 44d displays more indicators associated with an interaction
event, however, given that this galaxy contains a strong bar, the
interpretation of its velocity field is difficult and in addition
no valuable rotation curve can be derived.  This strong bar could
have been induced by past strong tidal interactions although a
secular origin could not be excluded. Gathering information for all
galaxies, we do not find signatures of ongoing merger. However, Verdes-Montenegro et al. (2001) 
classified this system as a ``Phase 3a'' group, i. e., most of the HI 
gas has been stripped from the disk of the galaxies, which is an evidence of 
galaxy-galaxy interaction. These findings are in agreement with the giant HI tail recently discovered in this compact group by Serra et al. (2013); the authors suggest that this tail could be formed by a tidal effect (caused by the group) over HCG 44d. We may speculate that these results can support the origin of the strong bar of HCG 44d.

We observed three galaxies in {\bf HCG 53}. Two of them show H$\alpha$
emission (53a, an SBbc and 53c, an SBd). The third one (53b) is an
S0 galaxy and does not have emission. HCG 53a does not have any
signatures of interaction.  HCG 53c has a few indicators flagged
positive. It has a misalignment between the optical and kinematic
major axis of the position angle.  This signature can be a result
of an interaction with its companion HCG 53b. In general, HCG 53
shows weak signatures of galaxy-galaxy interactions, being, therefore,
in an early stage of evolution..

We observed seven members of {\bf HCG 57} and four of them are
spiral galaxies, mainly galaxies a, b, d and h have types Sb, SBb,
SBc and SBb respectively. We detect clear H$\alpha$ emission only for member HCG 57d (a diffuse emission was detected in the central region of HCG 57a). HCG 57d has a regular velocity field, however, there is change in the kinematical position angle along the major axis. Both sides of its rotation curve do not match in the outer parts of this object and its velocity field seems to be connected with HCG 57a. A strong burst of star formation can be seen in a spiral arm. These interaction
indicators suggest that this object is in an interaction process,
and it is probably interacting with its companion galaxy HCG 57a. The lack of H$\alpha$ emission in other members of this group can be associated with the detection limit of our observations, given that this system was observed with poor seeing conditions. This is further discussed in the next subsection.

We observed members a, c and d of {\bf HCG 61} which have morphological
types S0a, Sbc and S0, respectively. We have detected weak and
non-extended H$\alpha$ emission in the centres of these galaxies
(in agreement with Vilchez et al. 1998, their Table 2, where the
H$\alpha$ emission for galaxies HCG 61a,c,d was classified as NE, i.e.,
nuclear emission). The morphological type of members a and c are in agreement with the lack of H$\alpha$ emission. However, for HCG 61c, an Sbc
galaxy, we would have expected a more pronounced and extended
gas disk. This could indicate that interactions have taken place
and have stripped the gas.

We observed four galaxies in {\bf HCG 69}, members a, b, c and
d, classified as Sc, SBb, S0 and SB0 respectively. We detected weak
H$\alpha$ emission in member HCG 69a, when we expected a strong
signal, given its morphological type, and we did not detect
emission in HCG 69b either, an SBb galaxy. This could be due to S/N
problem of the data or, if true in nature, could be due to interactions
which caused depletion of warm gas in the centres of the member galaxies.

We observed two galaxies of {\bf VV 304} and both display clear
signatures of galaxy-galaxy interactions. VV 304a has distorted
spiral arms, but the velocity field is not strongly perturbed.
However, we detect a disagreement between both sides of the rotation
curve. In the case of VV 304b, seven out of ten interaction indicators
are present. However, neither VV 304a nor VV 304b display optical tidal tails, which are 
clear signatures of strong galaxy encounters. This information suggests that the interaction process
in VV 304 is still mild. However, given the apparent proximity of
the members in VV 304, a merger event should occur.

The three observed galaxies in {\bf LGG 455} were detected in
H$\alpha$.  NGC 7232 shows several signatures of interaction, like
highly perturbed velocity field and disagreement between both sides
of the rotation curve. The same happens for NGC 7232B. NGC 7233
shows a perturbed velocity field, however, the rotation curve is
symmetric, which is obviously a result of the azimuthal averaging.

The three galaxies observed in {\bf Arp 314} were detected in
H$\alpha$.  The three members have most of the interaction indicators
flagged positive.  This fact shows that Arp 314 is in an advanced
stage of evolution. This is consistent with the HI envelope that
encloses this group.

\subsection {Lack of warm gas in Hickson group galaxies?}

Despite the fact that the compact groups studied in this paper are
rather close to us, we detect five groups that contain Sb or
later-type galaxies and which present weak or no H$\alpha$ emission.
These are groups HCG 1, HCG 14, HCG 57, HCG 61 and HCG 69.

For HCG 1ab, the H$\alpha$ emission is mainly concentrated in a bridge
between these two galaxies. In the case of HCG 61c, the
H$\alpha$ emission is faint and centrally concentrated (as also reported by H$\alpha$
imaging from Vilchez et al. 1998). In the case of HCG 69a the
emission is extended but quite faint.  The worst cases are for HCG
14ac and 57abd, where we do not detect any H$\alpha$ emission at
all. On the other hand, we detect groups having galaxies (Sb or
later) that display clear H$\alpha$ emission, like HCG 25a, 44d,
53c, VV 304a, b and Arp 314 NED 01, 02 and 03.  As mentioned, the
complexity of the H$\alpha$ emission and the lack of it in some
galaxies can be used as an interaction indicator.  However, we can
not rule out the possibility that, at least in a few of the cases
above, the lack of H$\alpha$ emission is linked to the low S/N of
the observations. In fact, this seems to be the case for HCG 57, given that Martinez et al. (2010) detected H$\alpha$ emission in galaxies where we do not detect any signal. This idea is consistent with the observing conditions on which were obtained the Fabry-Perot data of HCG 57 (under poor seeing conditions).

Assuming the lack of H$\alpha$ emission in some of the late type galaxies is real, we
suggest that this can be related with interactions and with a late
evolutionary stage of these compact groups. It is well known that
interactions  can remove some {\it neutral} gas of the galaxy disk
and this gas can be ejected into the intergalactic medium, which
has been shown by Verdes-Montenegro et al. (2001) to be at work in
several compact groups. This scenario can result in a lack of ionized
gas in the main body of the HCG galaxies. In fact, Plana et al.
(1999) studied the HCG 92, also known as the Stefan Quintet, and
they found that the spiral galaxies of this system do not show
H$\alpha$ emission or show little warm gas emission in their centres.
Taking into account all information gathered in this paper, we argue
this could indeed be the case for galaxies HCG 1ab and HCG 61c.  These
galaxies could have had their gaseous reservoirs removed, resulting
in a quenching of their star formation. In the case of HCG 1, most of the H$\alpha$ emission comes from a region located outside the main body of the galaxies. Note that groups HCG 1
and HCG 61 were found to have a normal HI content by Verdes-Montenegro
et al. (2001), when the whole group was taken into account, which
does not, however, conflict with the observations above.

Regarding the three other groups HCG 14, 57 and 69, we do not have enough data to conclude anything about their evolutionary stages. We note that these groups (as well as HCG 1, 25, 44 and 61) were observed with the OHP 1.93m-telescope, and not with the ESO 3.6m telescope where all the remaining data for compact group galaxies from this and our previous papers were obtained. One could argue that this may hint that the non-detection or weak detection is due to the lower S/N of the data. This idea is consistent with the study developed by Martinez et al. (2010), who found H$\alpha$ emission in most of the galaxies belonging to HGC 57. Therefore the conclusion is that we really need more kinematic data for groups HCG 14, 57 and 69, which we plan to collect in the future to settle this question. However, taking into account that the GHASP (OHP) and CIGALE (ESO) instruments are similar and the setups almost identical, both detectors are similar and have the same size,  the larger mirror size of the ESO telescope is almost compensated by the larger pixel size of the GHASP instrument. Thus, for a given extended source, observing time and atmospheric, telescope and filters transmissions, the SNR per pixel (but not by arcsec$^2$) should be roughly the same. Nevertheless, the main difference comes from the seeing conditions that dilute the H$\alpha$ emission; the average seeing at OHP being worse than that at ESO (see Table \ref{observinglog}), the SNR per pixel is on average higher for the ESO than for the OHP data. In addition, measurements of the narrow band filter transmission done after the OHP run, indicated that their transmission range between 70 and 80\% (which are the expected values) except for four of them: the filters used to observe HCG 1, 14, 69 and 93 drop more or less drastically down to 40, 25, 55 and 25\% respectively.

In this sample, we observe a total of 42 galaxies located in 12
compact groups. A total of 16/42 galaxies are E and S0/S0a. Three galaxies classified as S0a and two S0 galaxies display H$\alpha$ emission. 
The other 11 E/S0/S0a galaxies do not present any H$\alpha$ emission. The remaining
26/42 objects are non barred or barred spiral galaxies ranging from
Sa to Im morphological types. We were able to derive velocity fields
based on their H$\alpha$ emission for 18/26 spiral galaxies, 5/16 E/S0/S0a galaxies and one Im galaxy. Among these 24 galaxies for which a velocity map was obtained, for 15 of them
we could derive a rotation curve and among these 15 rotation curves
only 2/15 show no signatures of interactions. In forthcoming papers, we will analyse
these results in a broader context, including previously published
results for other compacts groups.

\section{acknowledgements}

We thank the anonymous referee for the useful comments that greatly improved this paper. We warmly thank Philippe Balard, Olivier Boissin, Beno\^it Epinat \& Jean-Luc Gach for making the setup of the GHASP instrument on the 1.93m at OHP and Jacques Boulesteix \& Jean-Luc Gach for the setup of the CIGALE instrument on the 3.6m at ESO. We also thank the OHP \&  ESO technical teams for support.  S. T-F acknowledges the financial support of the project \mbox{CONICYT PAI/ACADEMIA 7912010004}. CB, MM and PA thank the PNCG (Program National Cosmologie et Galaxies) for partial funding of this project. CMdO, PA, CB and S.T-F would like to thank LabCosmos and USP-Cofecub for funding of collaboration visits in France, Brazil and Chile. CMdO acknowledges funding from FAPESP (grant 2006/56213-9) and CNPq. HP thanks CNPq/CAPES for its financial support using the PROCAD project 552236/2011-0. S. T-F and D. O. acknowledges the financial support of the Direcci\'on de Investigaci\'on of the Universidad de La Serena, through a ``Concurso de Apoyo a Tesis 2013'', under contract PT13145. This research has made use of the NASA/IPAC Extragalactic Database (NED) which is operated by the Jet Propulsion Laboratory, California Institute of Technology, under contract with the National Aeronautics and Space Administration. We also acknowledge the use of the HyperLeda database (http://leda.univ-lyon1.fr).

\clearpage

\appendix

\section{Digital sky survey optical images, H$\alpha$ monochromatic maps, velocity fields, velocity dispersion maps and rotation curves for the galaxies of this sample.}
\label{appendixa}

\begin{figure*}
\includegraphics[width=\textwidth]{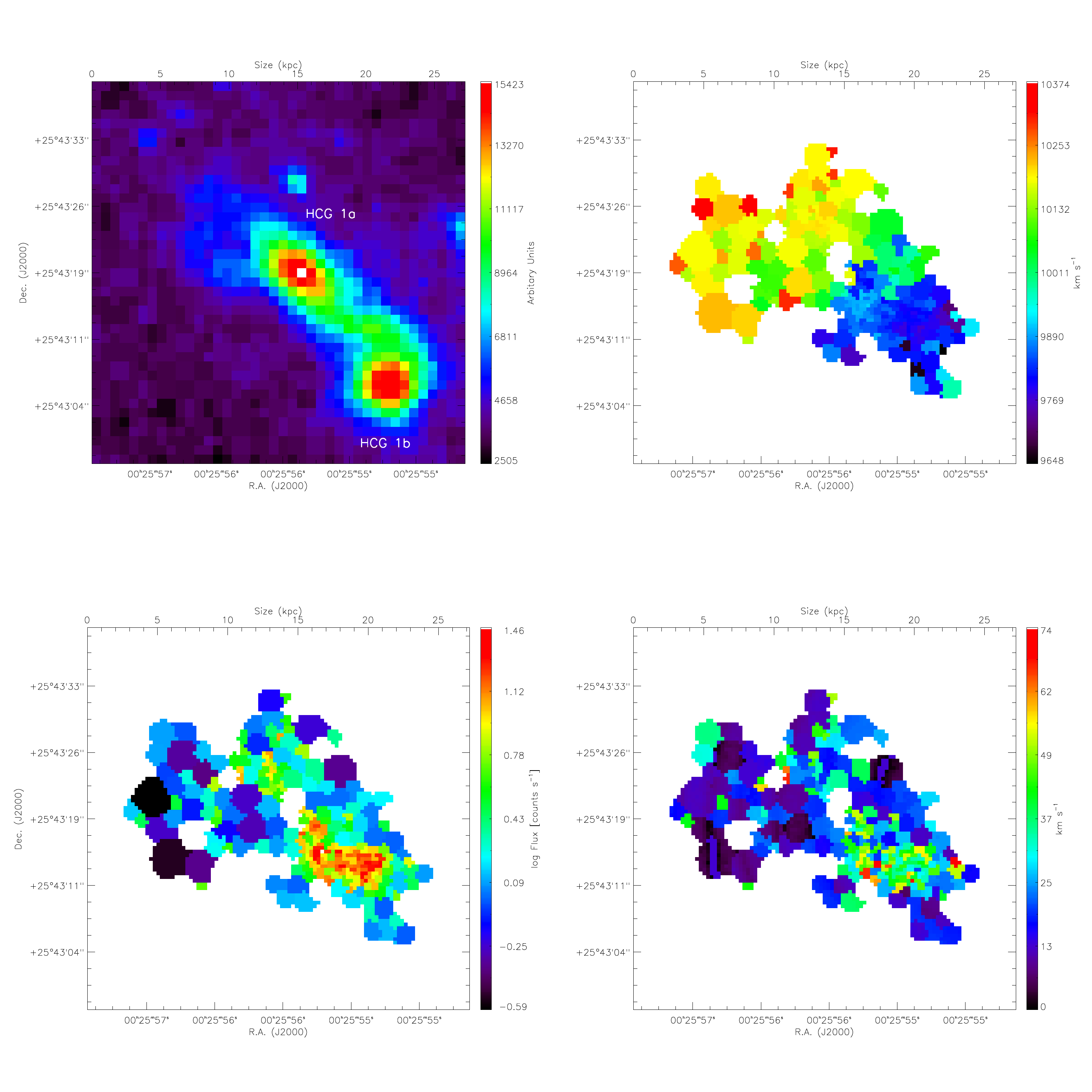}
\caption{Maps for HCG 1ab. Top left: B band image from DSS. Top right: velocity field. Bottom left: monochromatic image. Bottom right: velocity dispersion map.}
\label{maps_HCG1}
\end{figure*}

\begin{figure*}
\includegraphics[width=\textwidth]{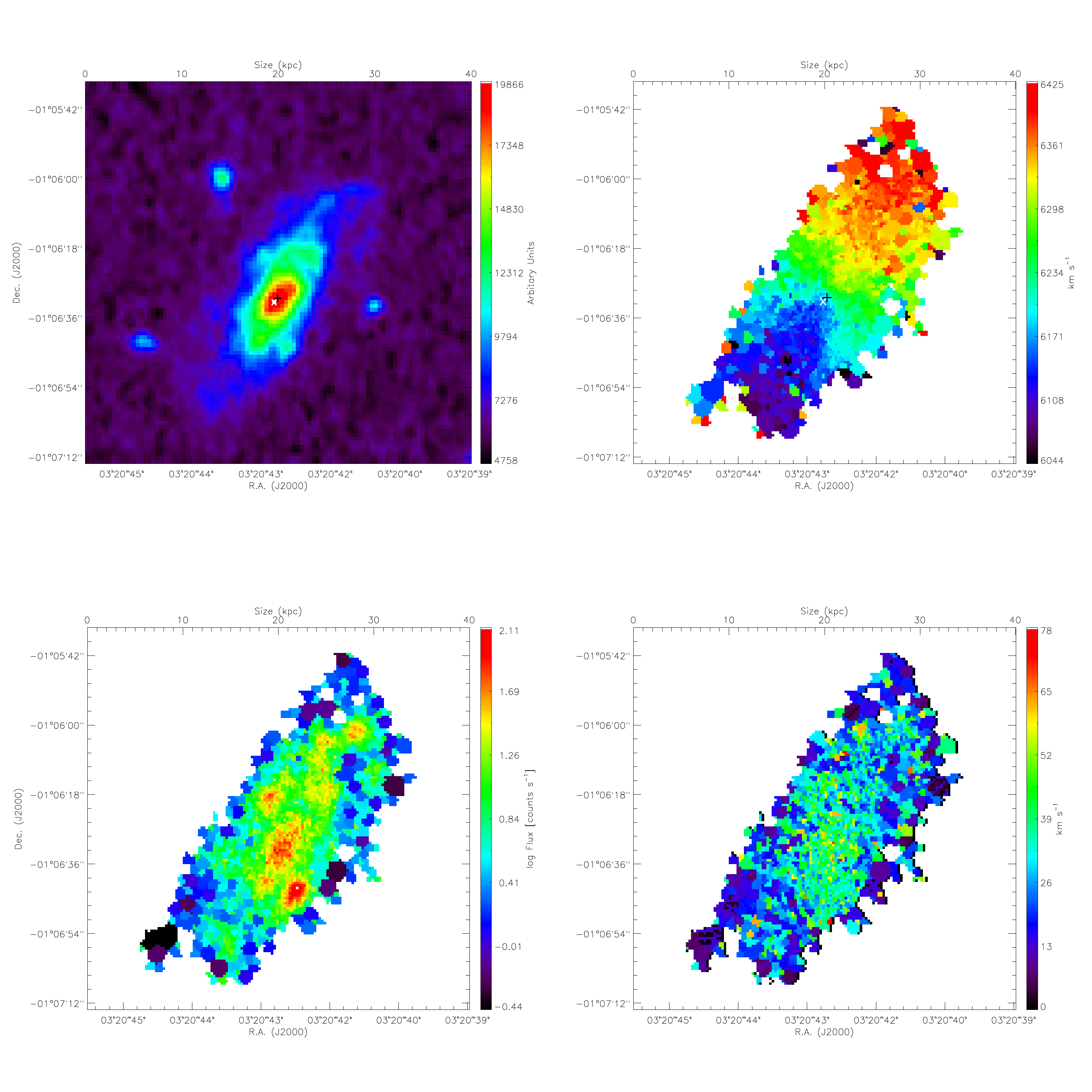}
\caption{Maps for HCG 25a. Top left: B band image from DSS. Top right: velocity field. Bottom left: monochromatic image. Bottom right: velocity dispersion map. The white ``x'' sign indicate the position of the photometric centre while the black ``+'' sign indicate the position of the kinematic centre.}
\label{maps_HCG25}
\end{figure*}

\clearpage

\begin{figure*}
\includegraphics[width=\textwidth]{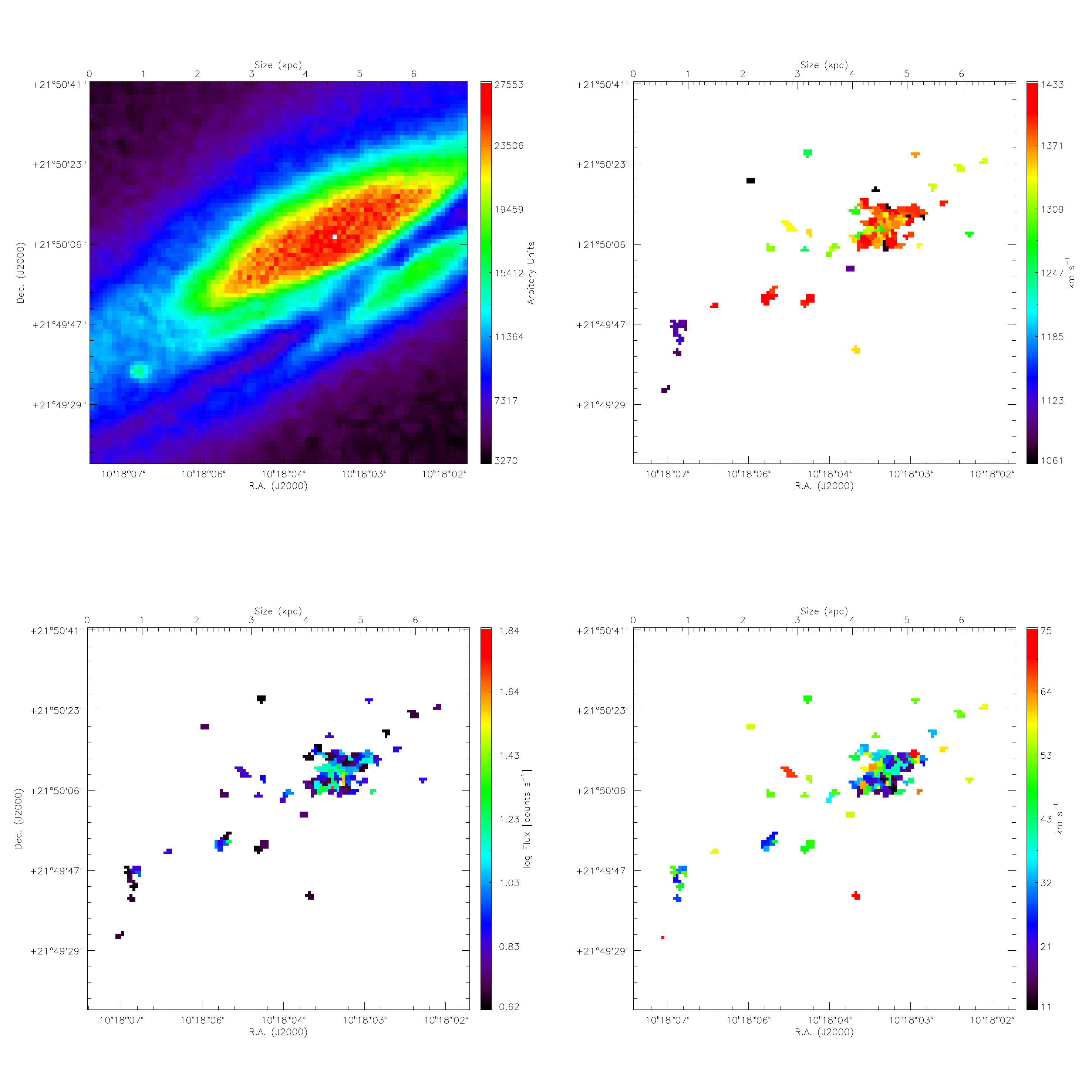}
\caption{Maps for HCG 44a. Top left: B band image from DSS. Top right: velocity field. Bottom left: monochromatic image. Bottom right: velocity dispersion map.}
\label{maps_HCG44a}
\end{figure*}

\begin{figure*}
\includegraphics[width=\textwidth]{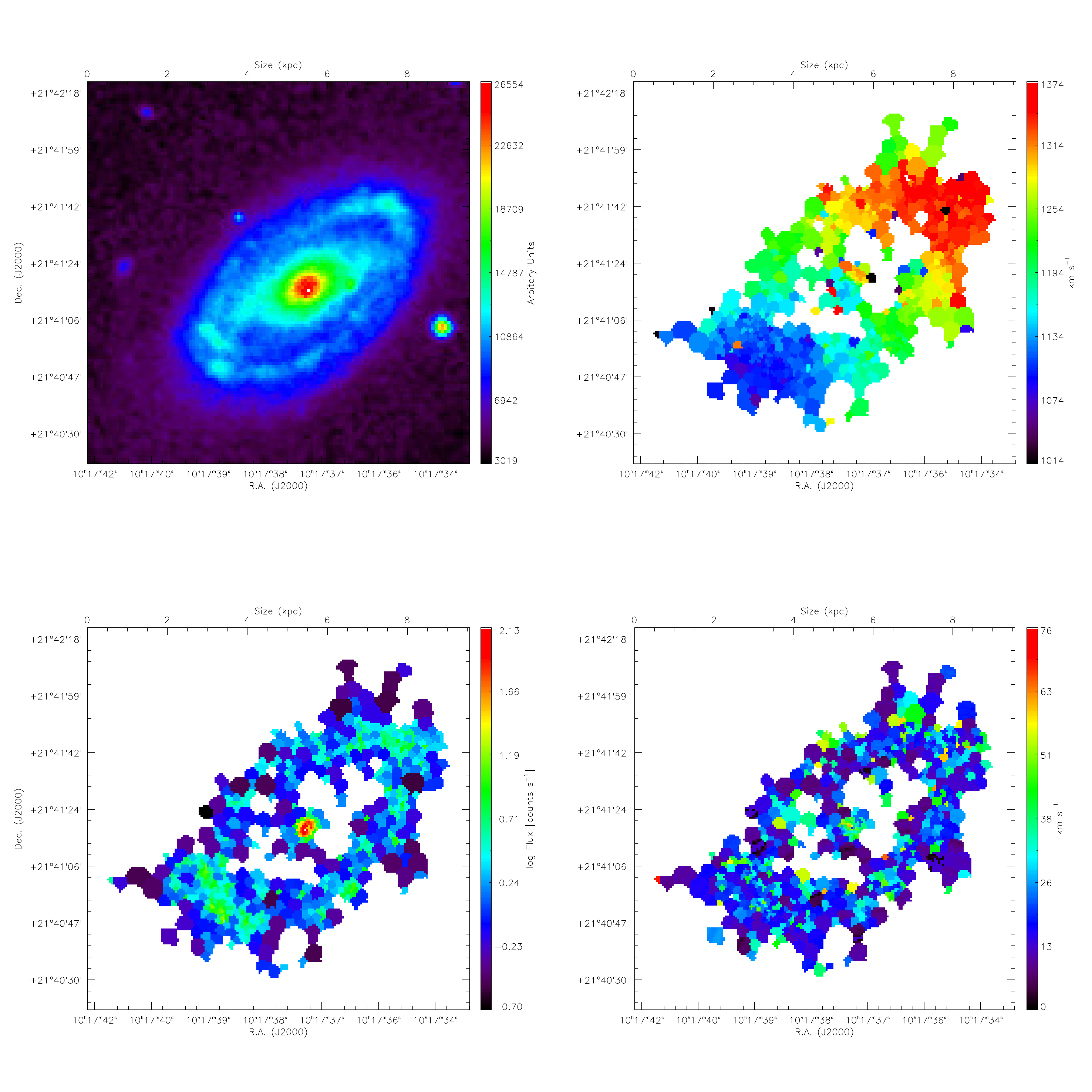}
\caption{Maps for HCG 44c. Top left: B band image from DSS. Top right: velocity field. Bottom left: monochromatic image. Bottom right: velocity dispersion map.}
\label{maps_HCG44c}
\end{figure*}

\begin{figure*}
\includegraphics[width=\textwidth]{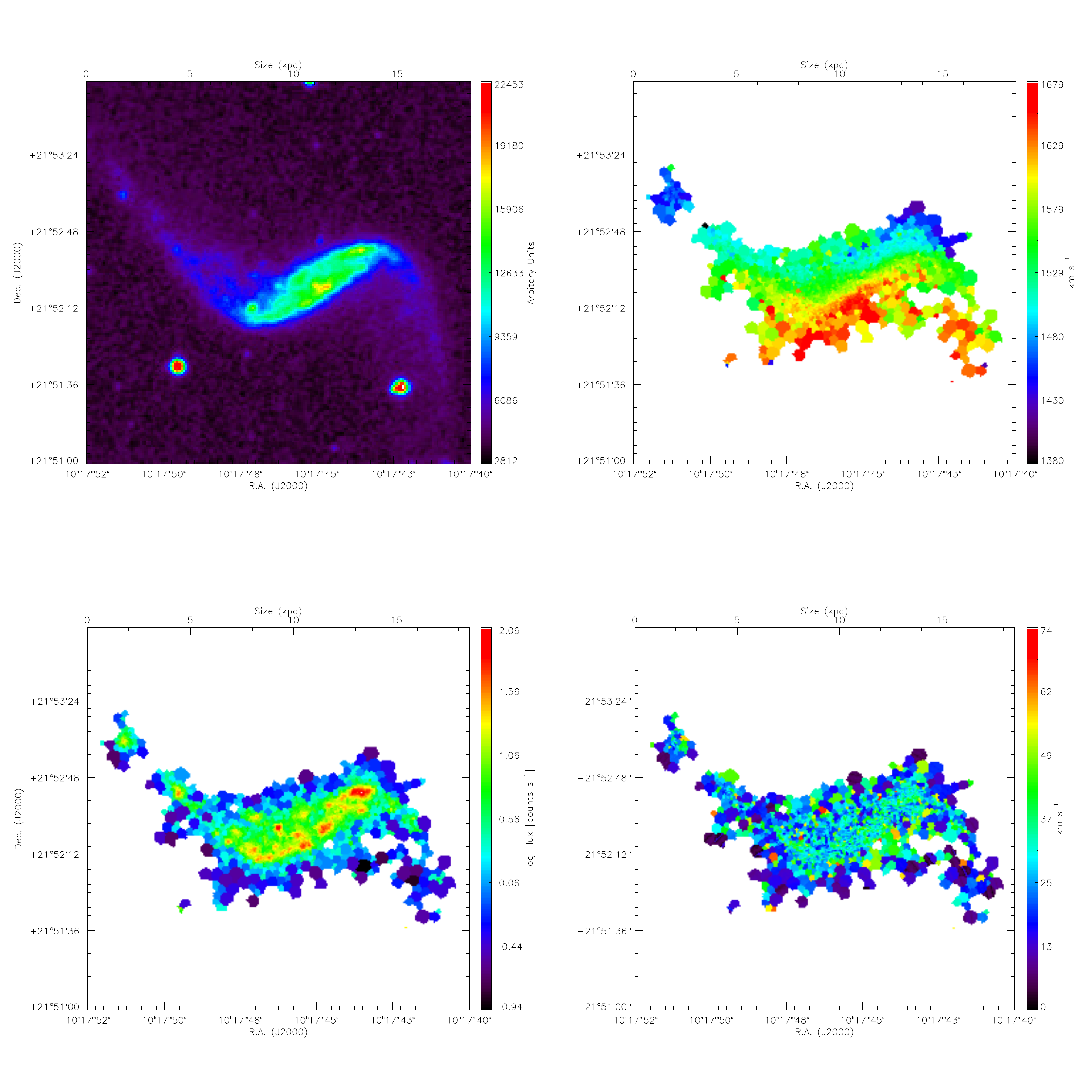}
\caption{Maps for HCG 44d. Top left: B band image from DSS. Top right: velocity field. Bottom left: monochromatic image. Bottom right: velocity dispersion map.}
\label{maps_HCG44d}
\end{figure*}

\clearpage

\begin{figure*}
\includegraphics[width=\textwidth]{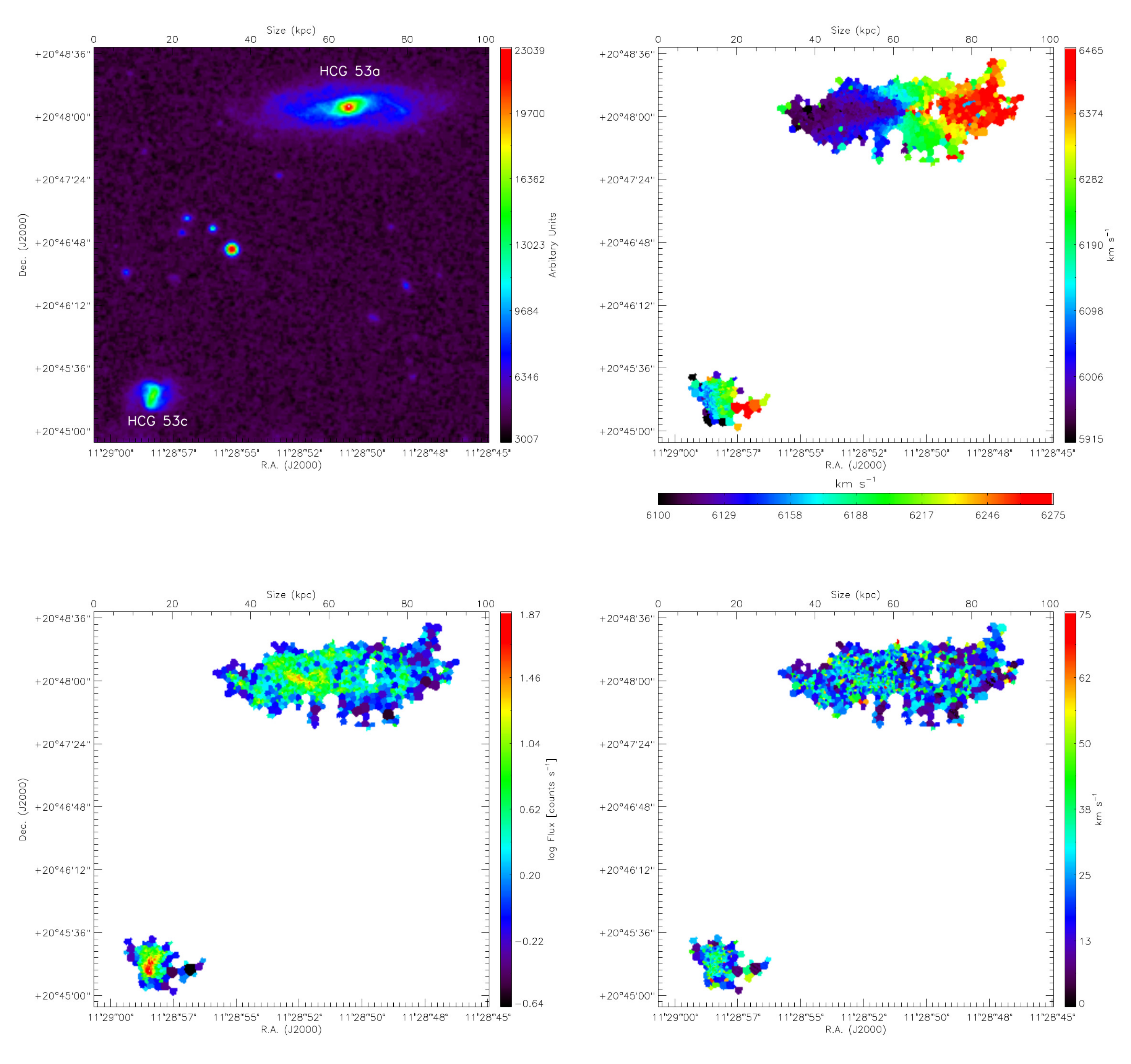}
\caption{Maps for HCG 53ac. Top left: B band image from DSS. Top right: velocity field, where the bottom colour bar corresponds to HCG 53c. Bottom left: monochromatic image. Bottom right: velocity dispersion map.}
\label{maps_HCG53abc}
\end{figure*}

\begin{figure*}
\includegraphics[width=\textwidth]{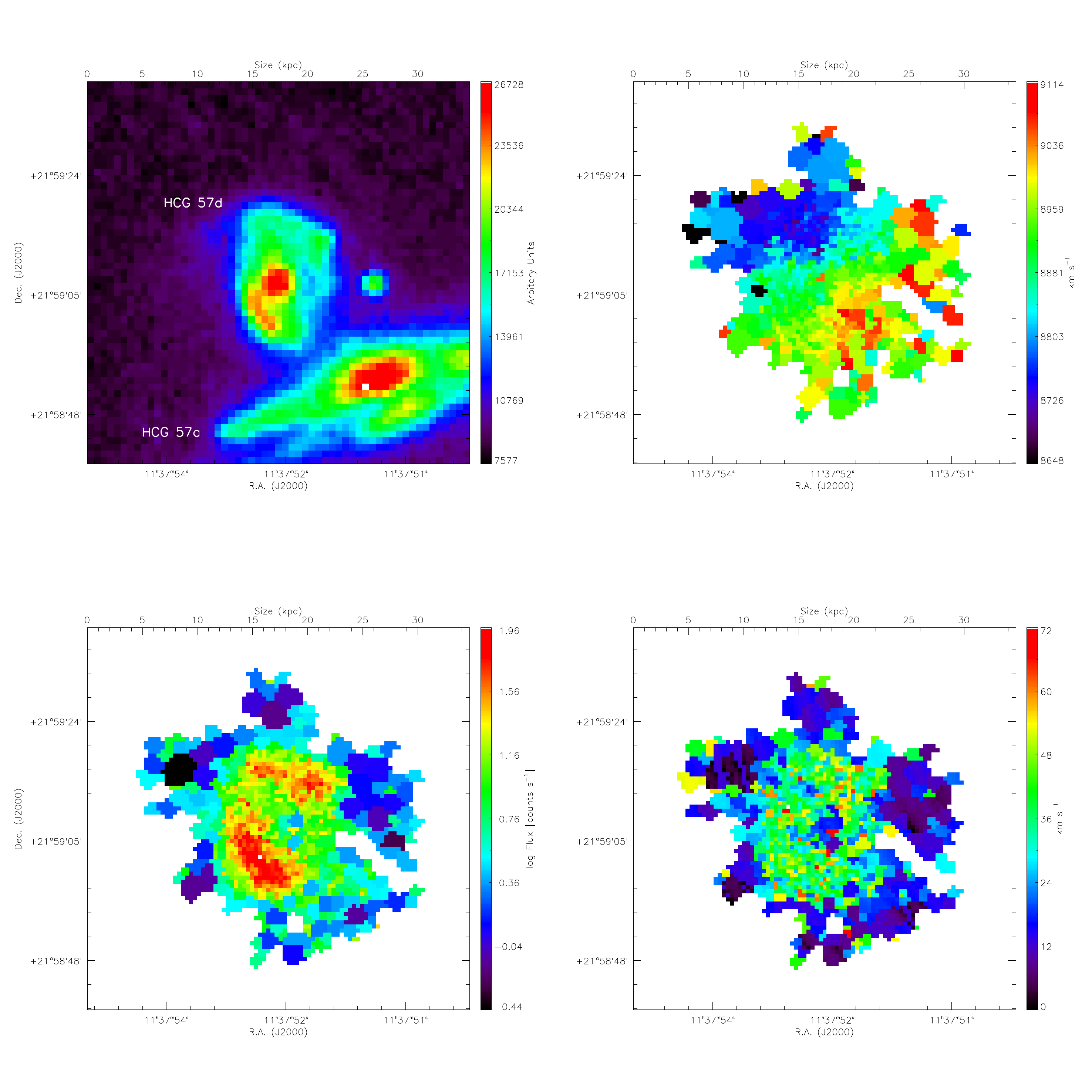}
\caption{Maps for HCG 57ad. Top left: B band image from DSS. Top right: velocity field. Bottom left: monochromatic image. Bottom right: velocity dispersion map.}
\label{maps_HCG57bdeg}
\end{figure*}

\clearpage

\begin{figure*}
\includegraphics[width=\textwidth]{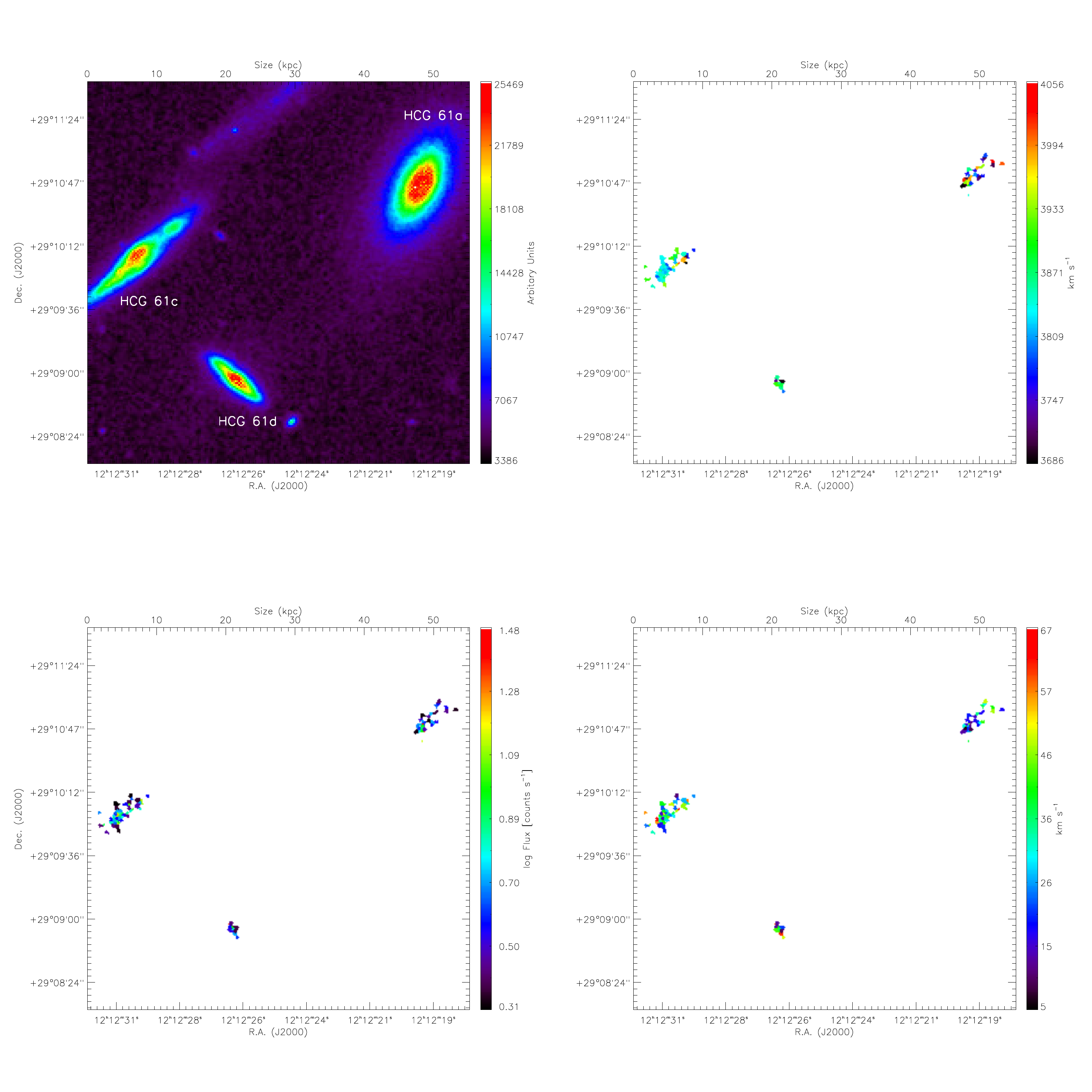}
\caption{Maps for HCG 61acd. Top left: B band image from DSS. Top right: velocity field. Bottom left: monochromatic image. Bottom right: velocity dispersion map.}
\label{maps_HCG61}
\end{figure*}

\begin{figure*}
\includegraphics[width=\textwidth]{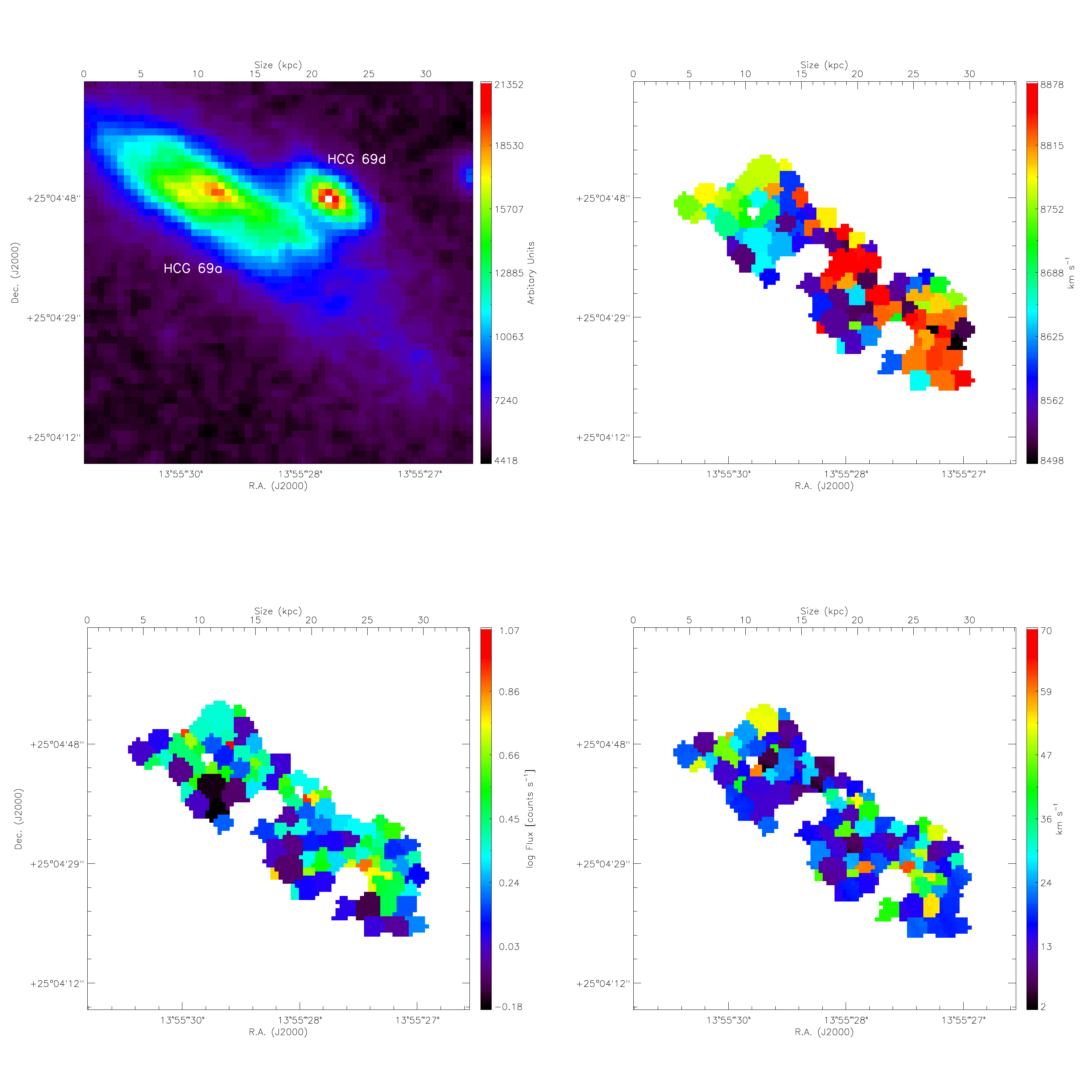}
\caption{Maps for HCG 69ad. Top left: B band image from DSS. Top right: velocity field. Bottom left: monochromatic image. Bottom right: velocity dispersion map.}
\label{maps_HCG69abc}
\end{figure*}

\clearpage

\begin{figure*}
\includegraphics[width=\textwidth]{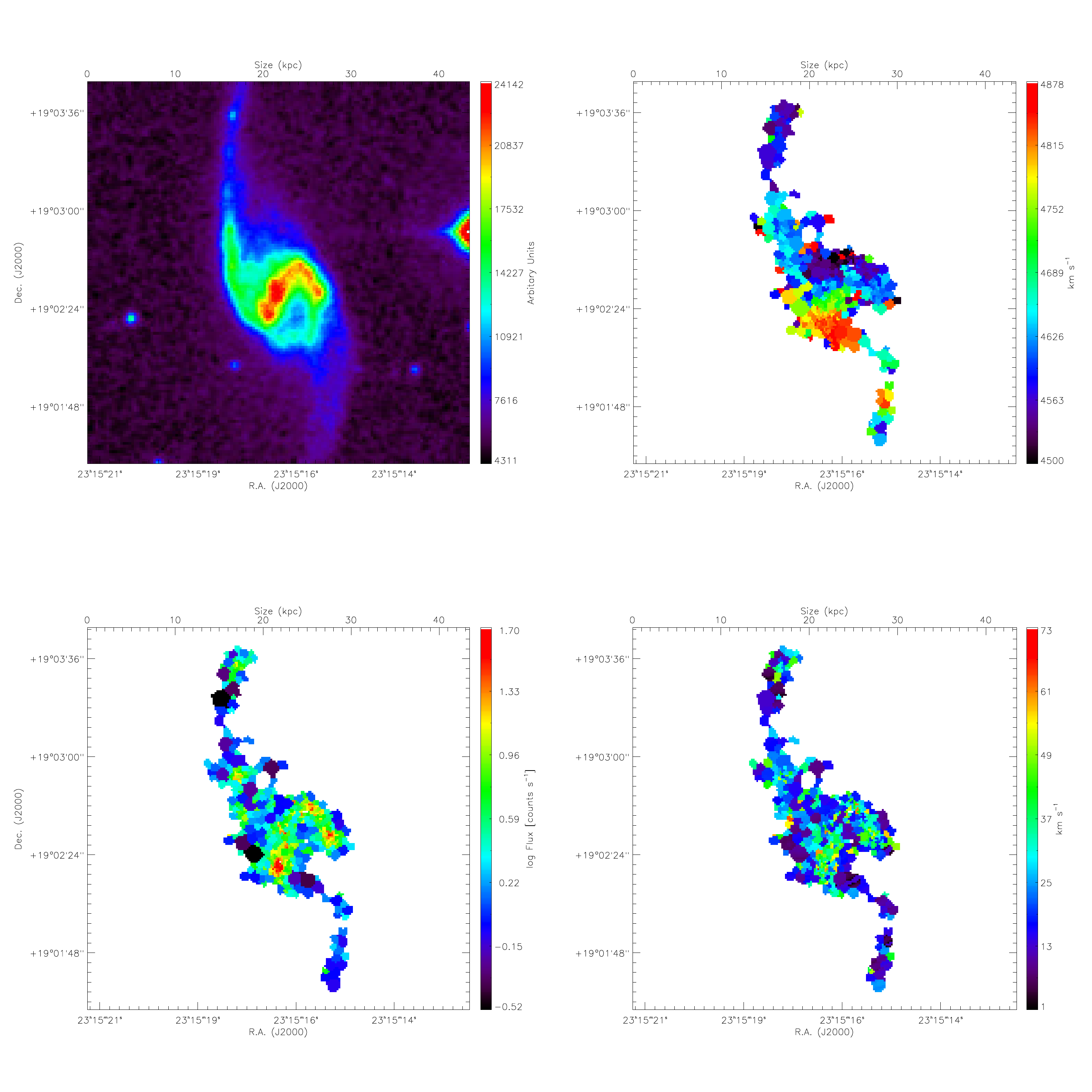}
\caption{Maps for HCG 93b. Top left: B band image from DSS. Top right: velocity field. Bottom left: monochromatic image. Bottom right: velocity dispersion map.}
\label{maps_HCG93b}
\end{figure*}

\begin{figure*}
\includegraphics[width=\textwidth]{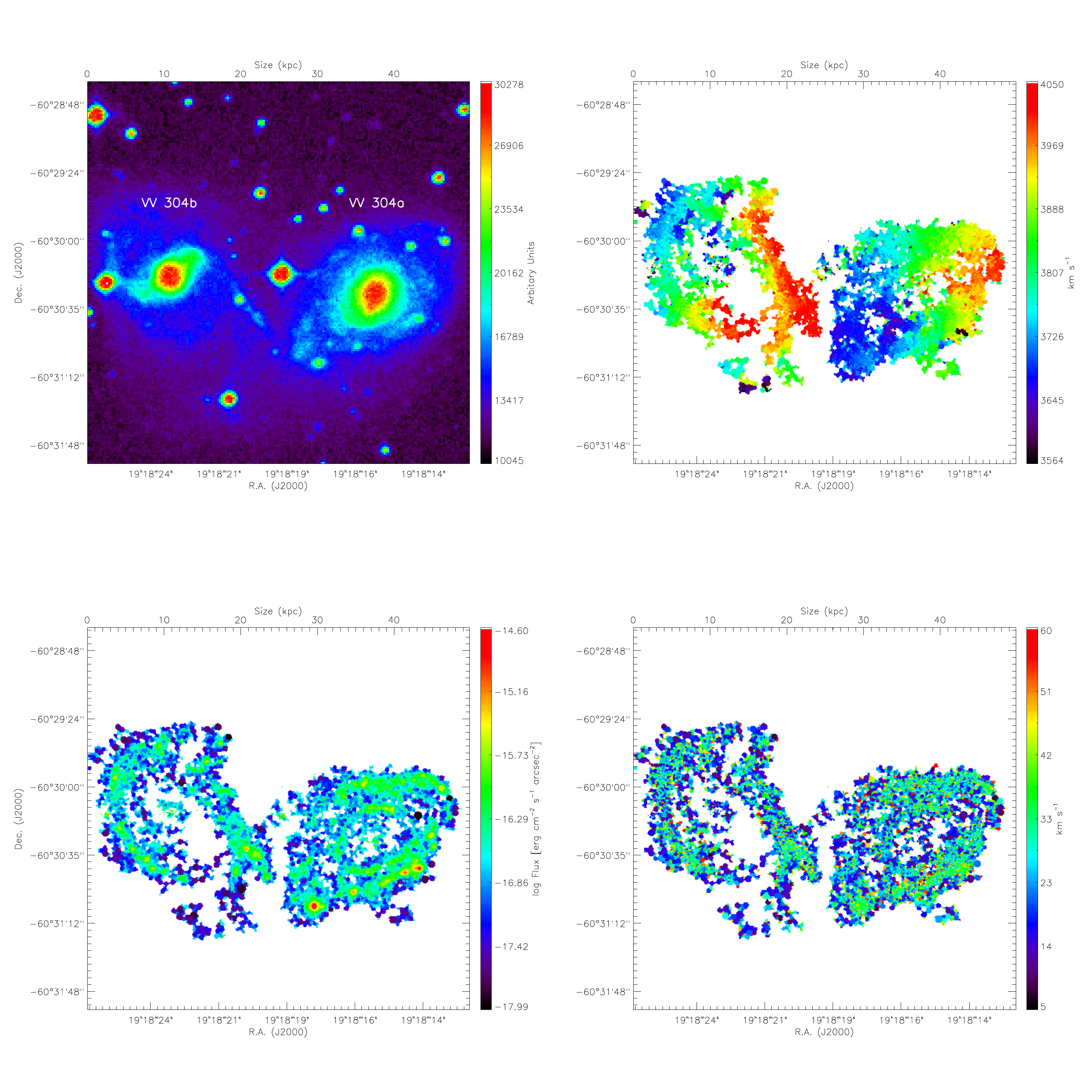}
\caption{Maps for VV 304ab. Top left: B band image from DSS. Top right: velocity field. Bottom left: monochromatic image. Bottom right: velocity dispersion map.}
\label{vfvv304_ned}
\end{figure*}

\begin{figure*}
\includegraphics[width=\textwidth]{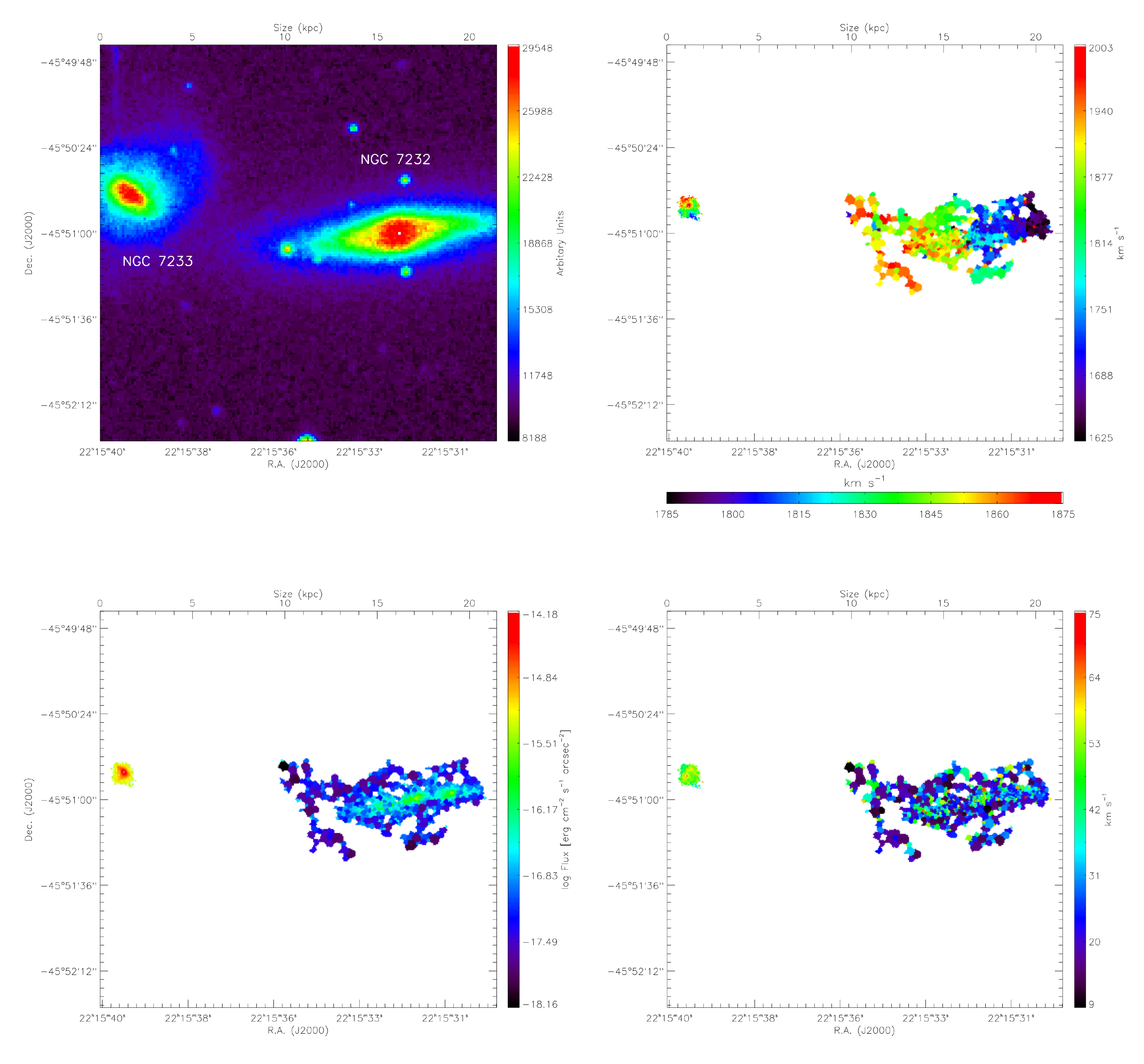}
\caption{Maps for LGG 455. Top left: B band image from DSS. Top right: velocity field, where the bottom colour bar corresponds to NGC 7233. Bottom left: monochromatic image. Bottom right: velocity dispersion map.}
\label{vflgg455ab}
\end{figure*}

\begin{figure*}
\includegraphics[width=\textwidth]{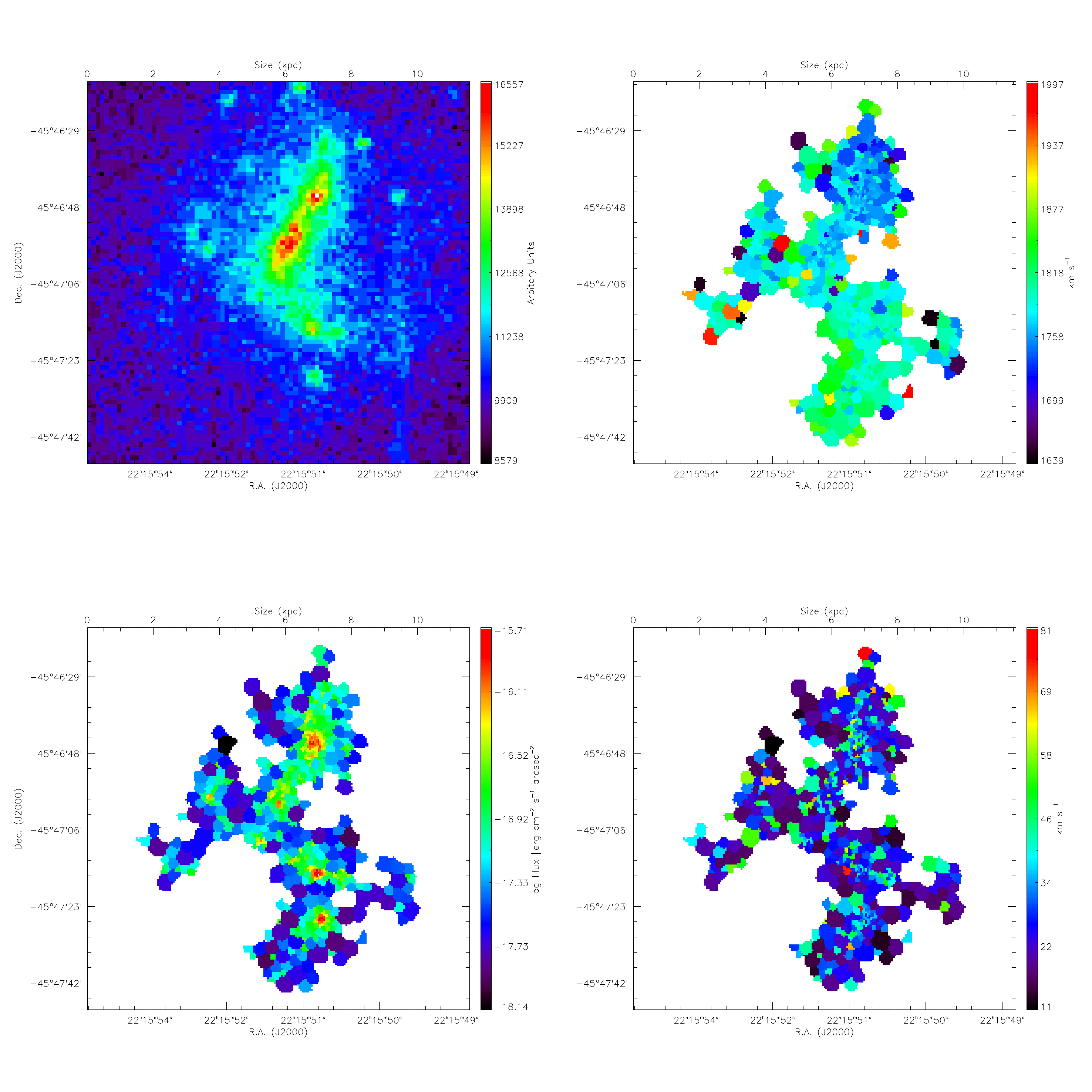}
\caption{Maps for LGG 455 (member NGC 7232B). Top left: B band image from DSS. Top right: velocity field. Bottom left: monochromatic image. Bottom right: velocity dispersion map.}
\label{vflgg455c}
\end{figure*}

\begin{figure*}
\includegraphics[width=\textwidth]{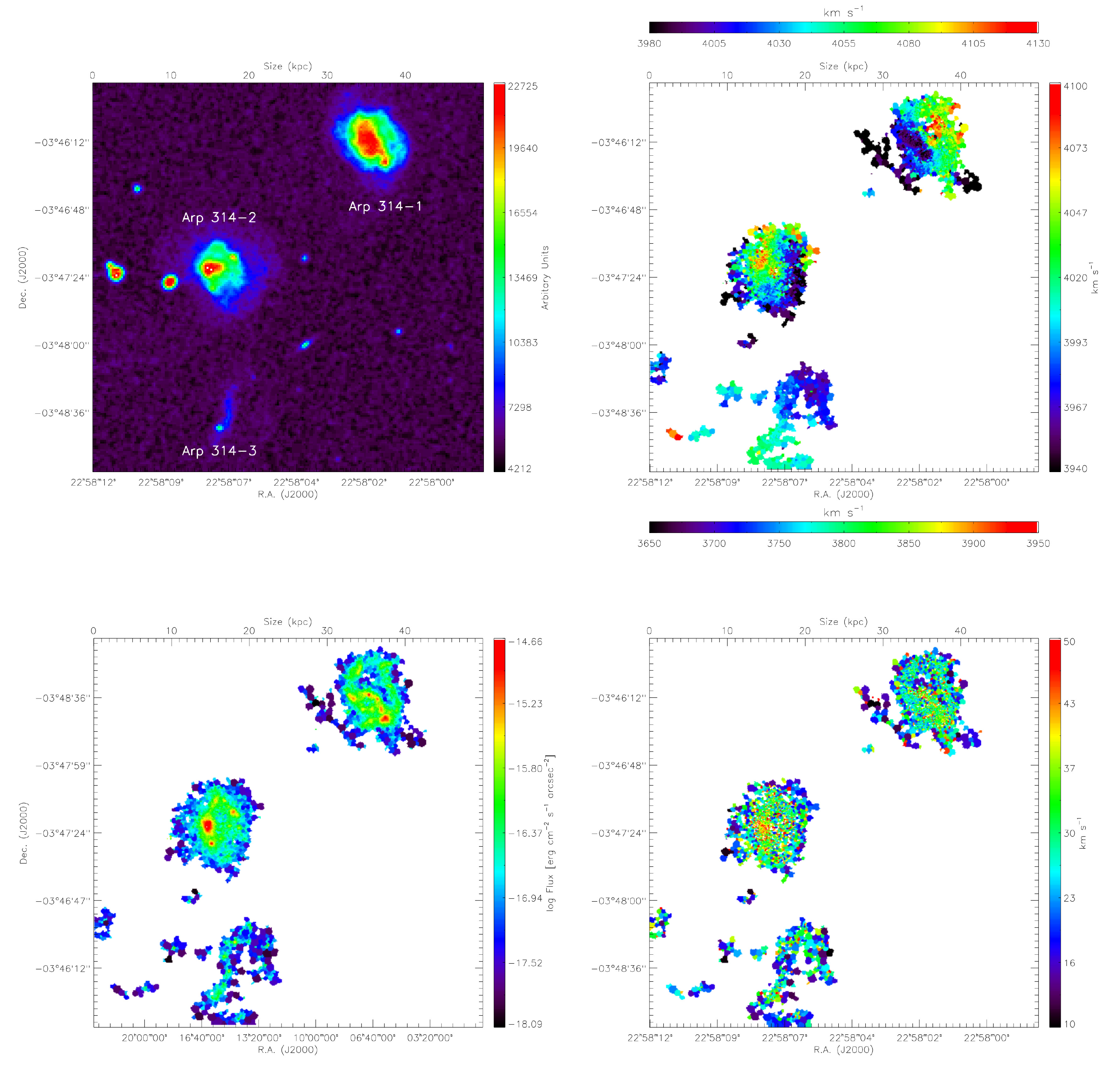}
\caption{Maps for Arp 314. Top left: B band image from DSS. Top right: velocity field, where the upper, right and bottom colour bars correspond to Arp 314-1, Arp 314-2 and Arp 314-3, respectively. Bottom left: monochromatic image. Bottom right: velocity dispersion map.}
\label{vfarp314}
\end{figure*}

\clearpage

\section{Rotation curves.}
\label{appendixb}

\begin{figure*}
\centering
\includegraphics[width=0.93\columnwidth]{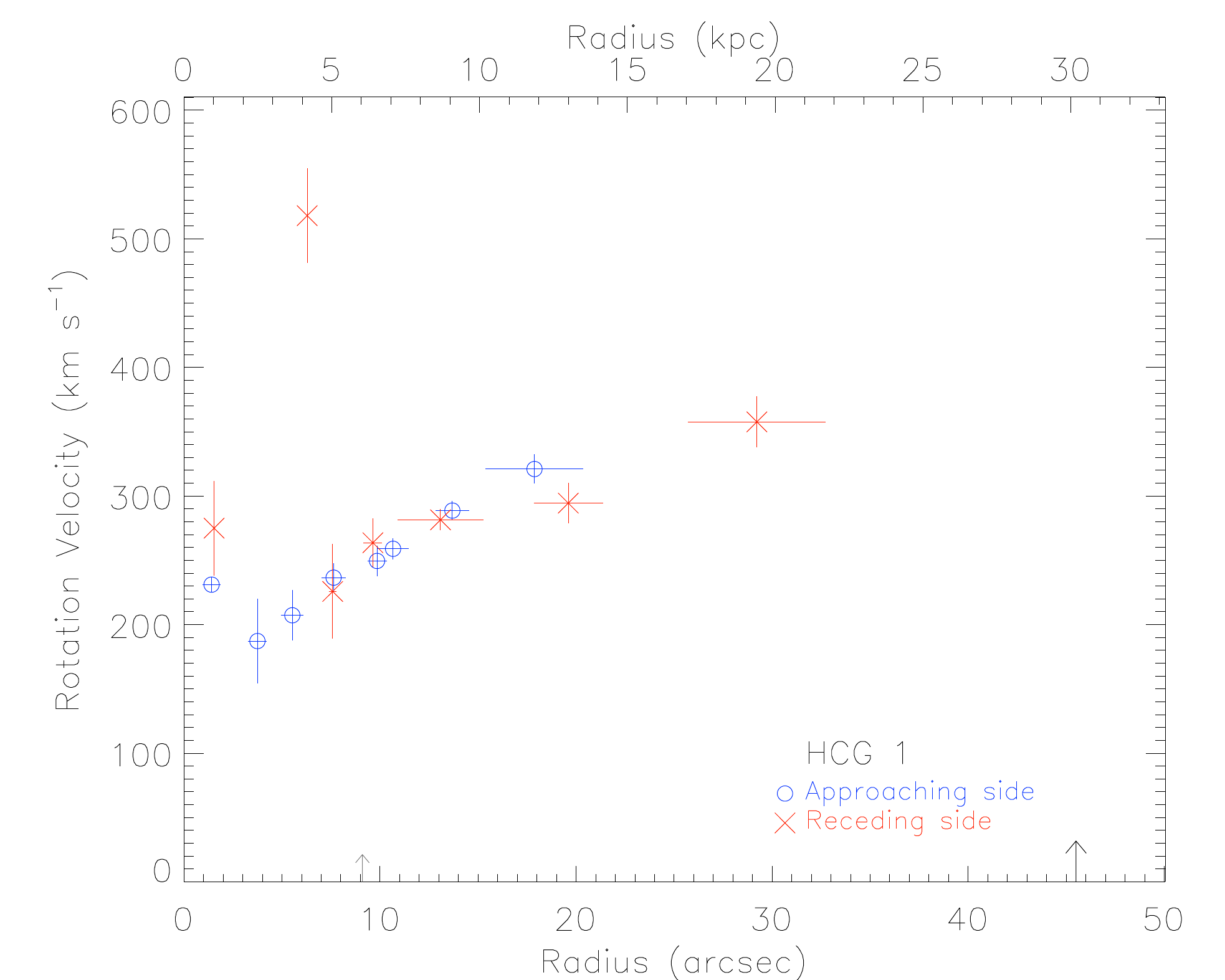}
\includegraphics[width=0.93\columnwidth]{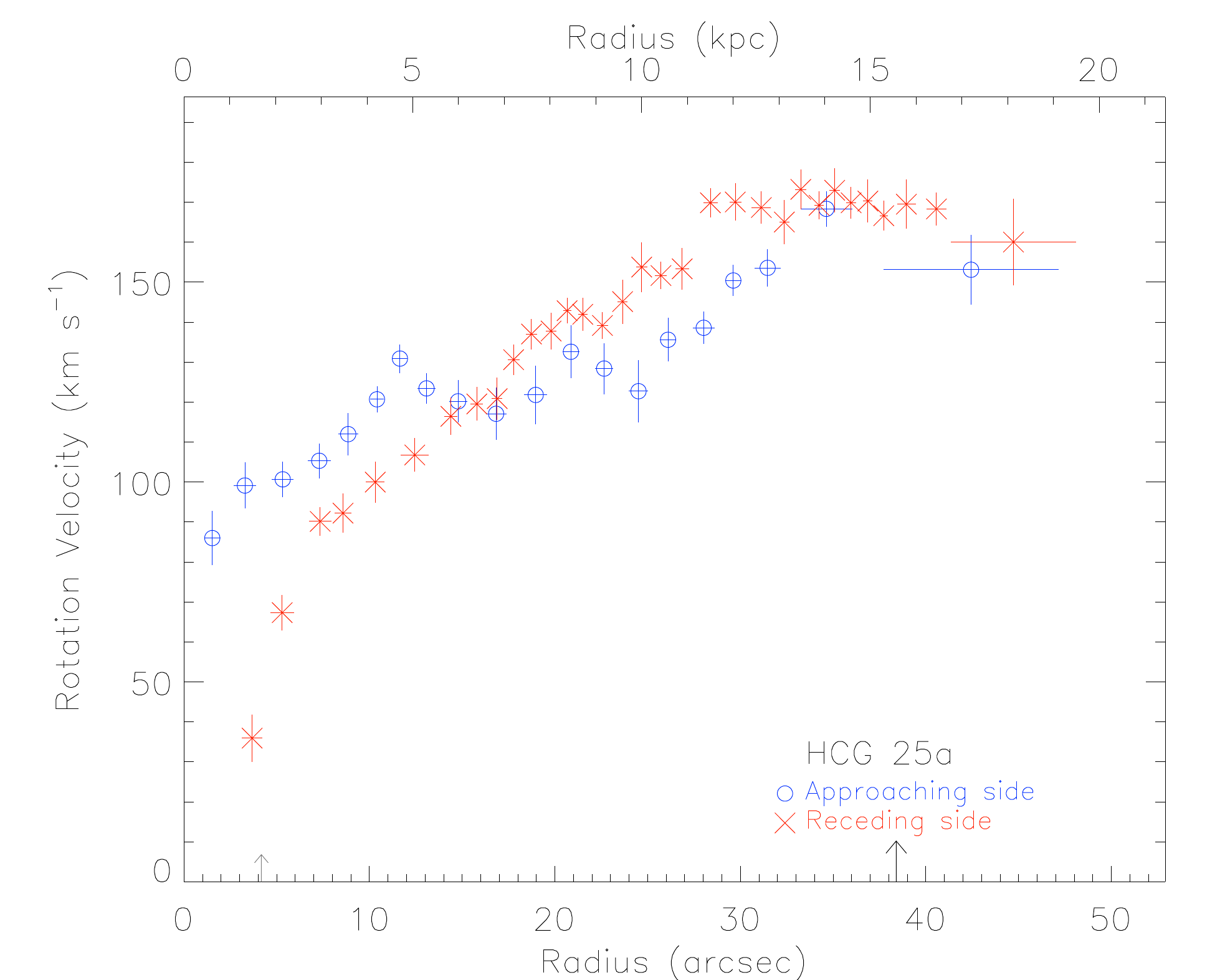}\\
\includegraphics[width=0.93\columnwidth]{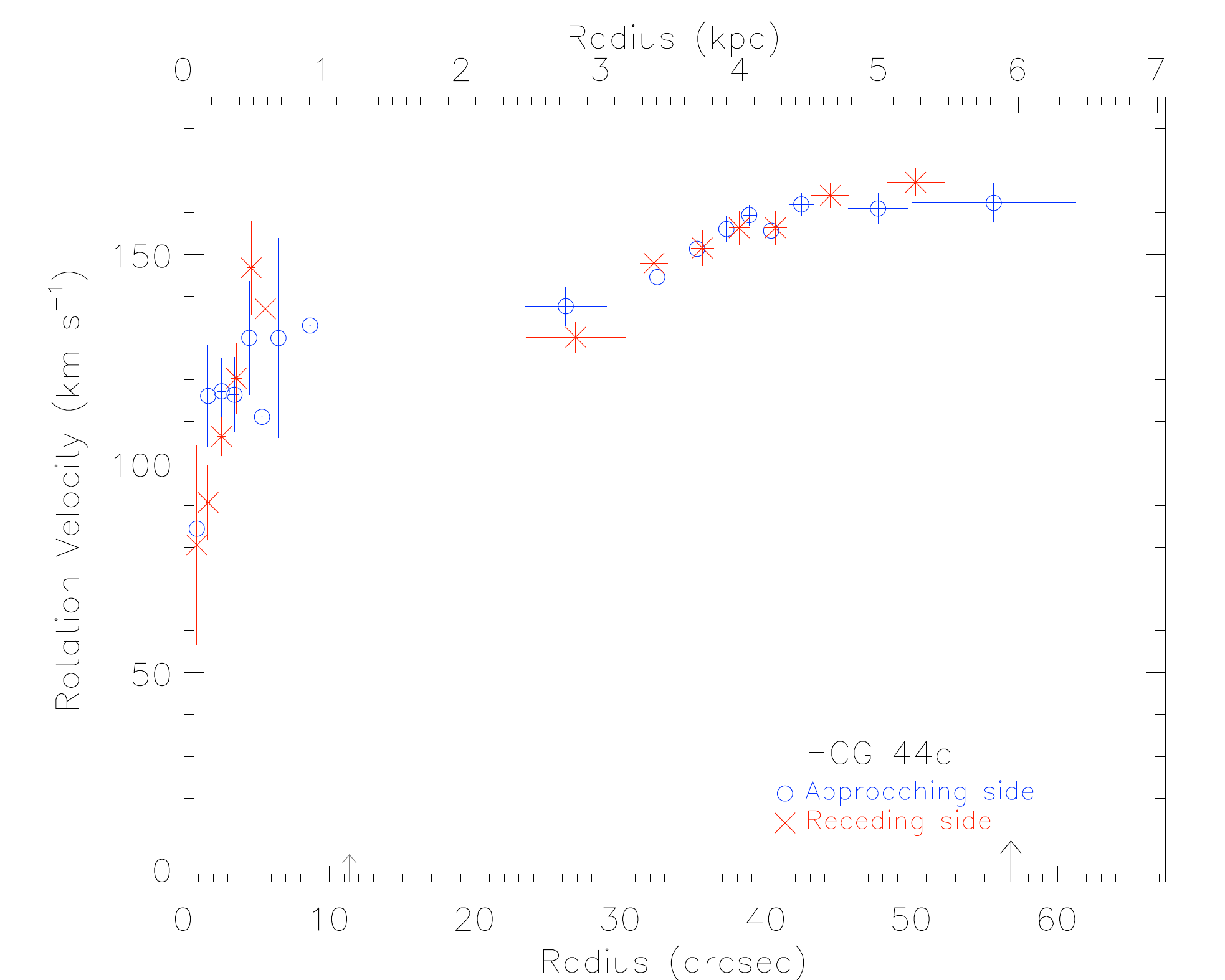}
\includegraphics[width=0.93\columnwidth]{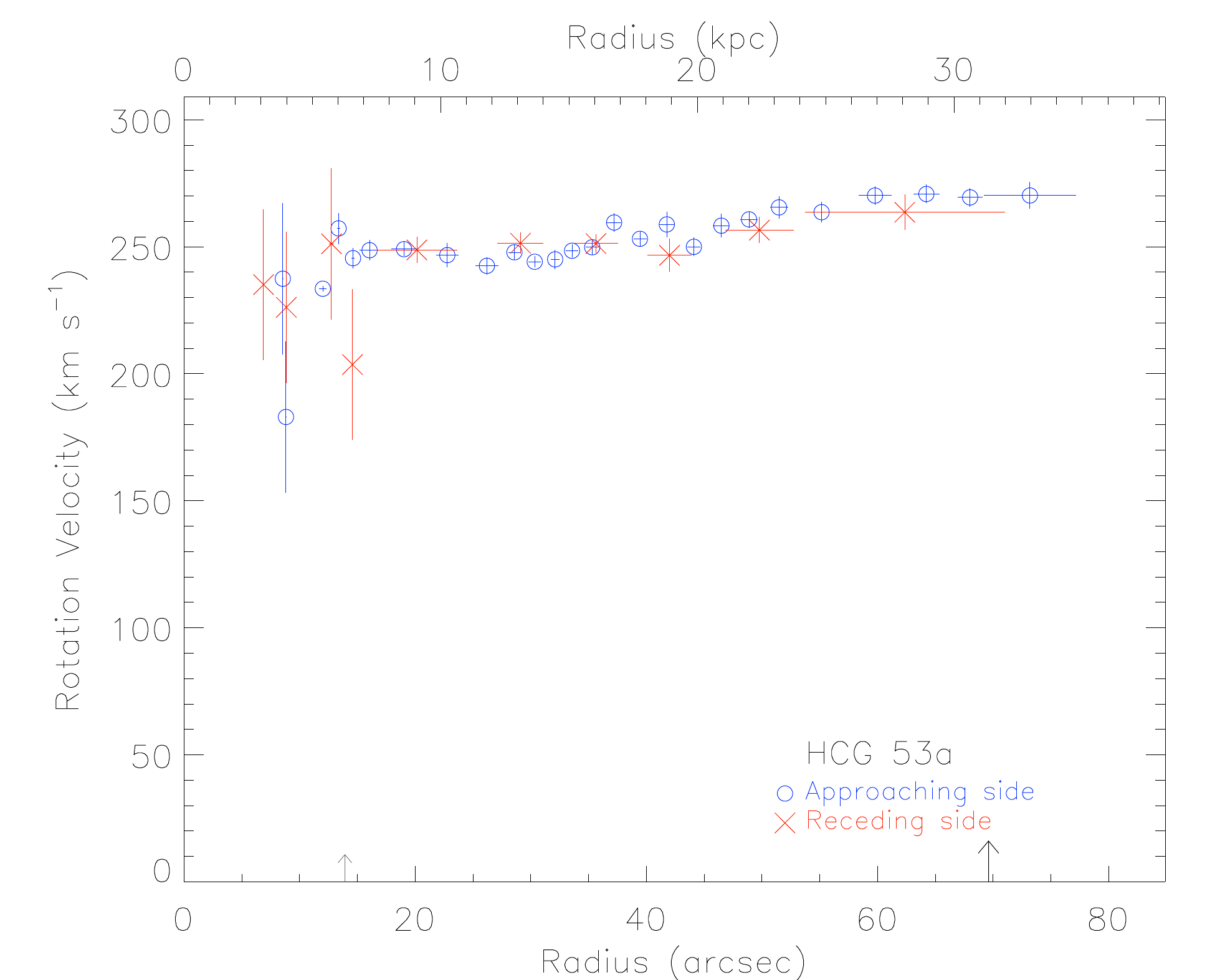}\\
\includegraphics[width=0.93\columnwidth]{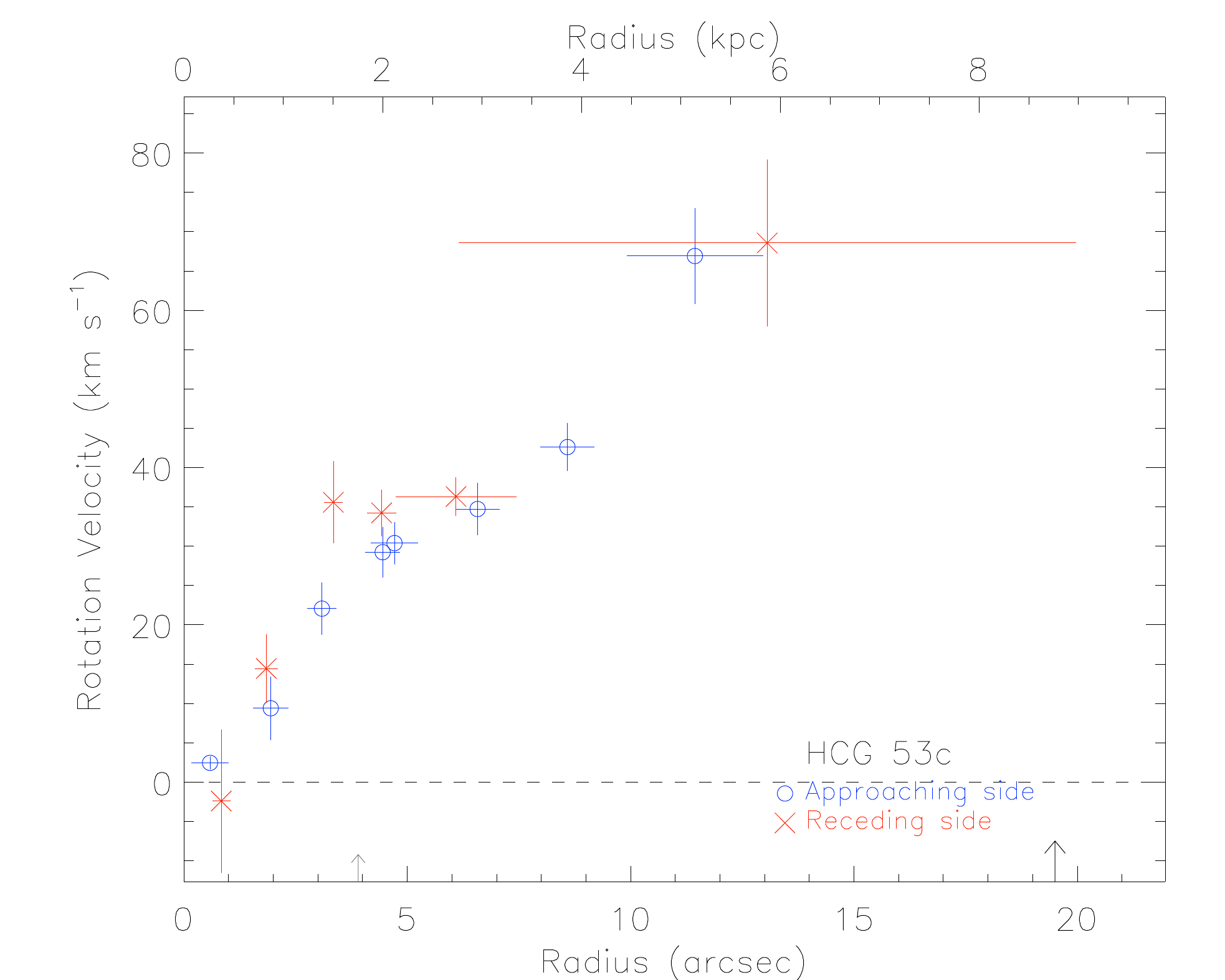}
\includegraphics[width=0.93\columnwidth]{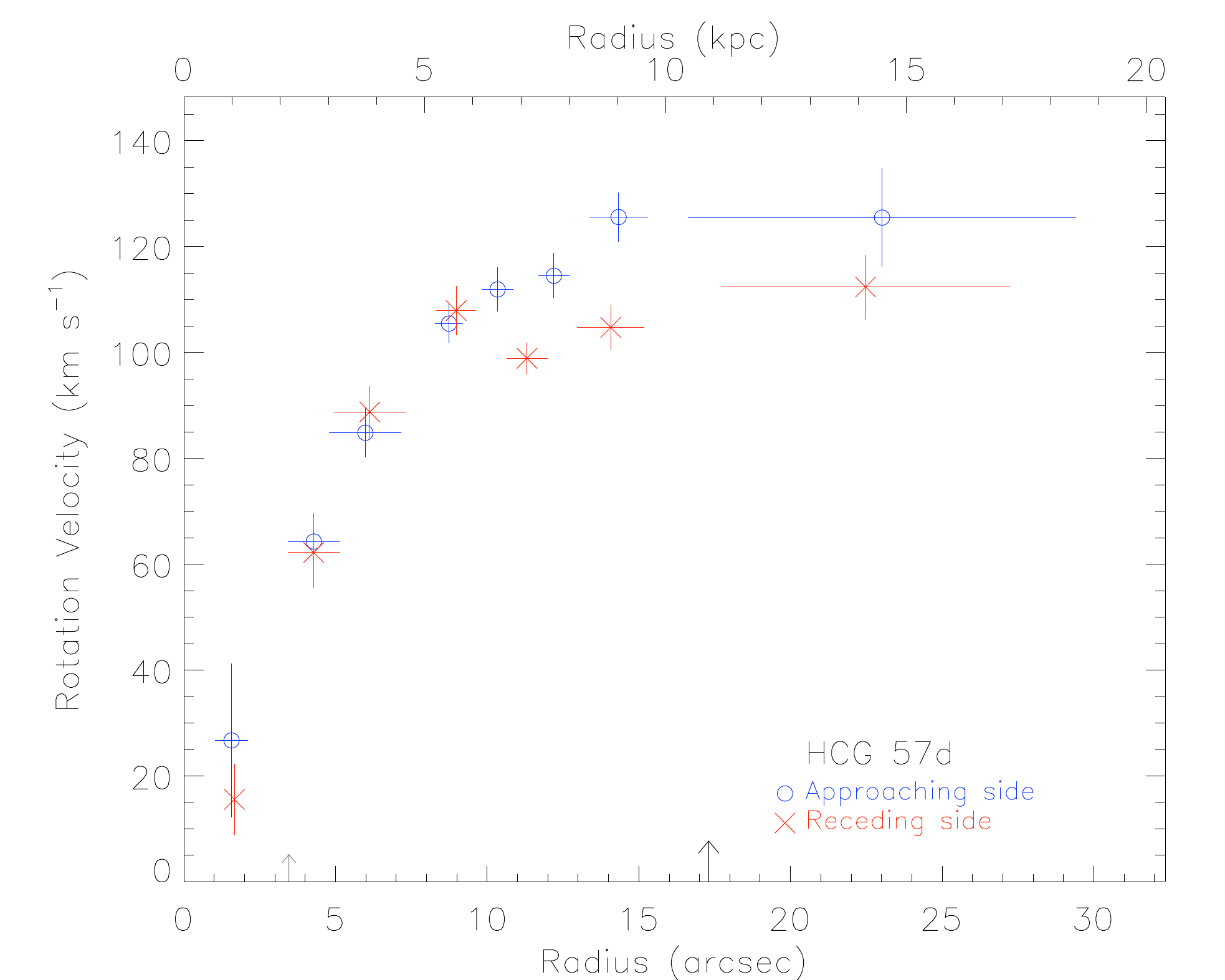}
\caption{Rotation curves in this sample. The PA and inclination were determinated automatically from the model. The centre was fixed (mophological centre). The black vertical arrow in the x-axis represents the radius R25 while the smaller grey arrow in the x-axis represents the transition radius that is defined by the first ring that contains more than 25 uncorrelated bins in the velocity field (see Epinat et al. 2008a). The horizontal error bars represent the width of the rings containing 25 bins except in the inner region (defined by the grey arrow). The vertical bars are the 1$\sigma$ uncertainty in the rotation velocity determination within each ring.}
\label{rc1}
\end{figure*}

\begin{figure*}
\ContinuedFloat
\centering
\includegraphics[width=0.93\columnwidth]{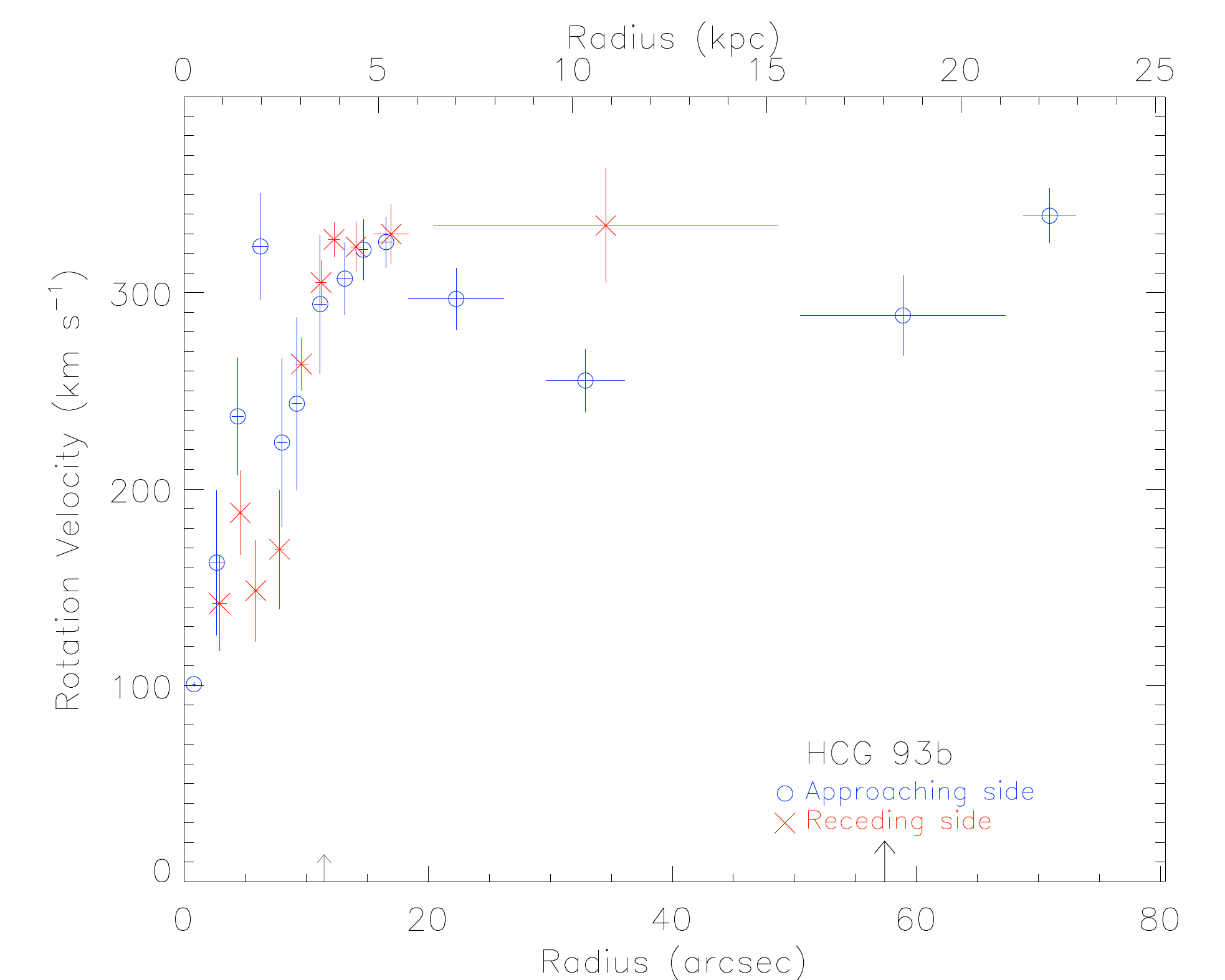}
\includegraphics[width=0.93\columnwidth]{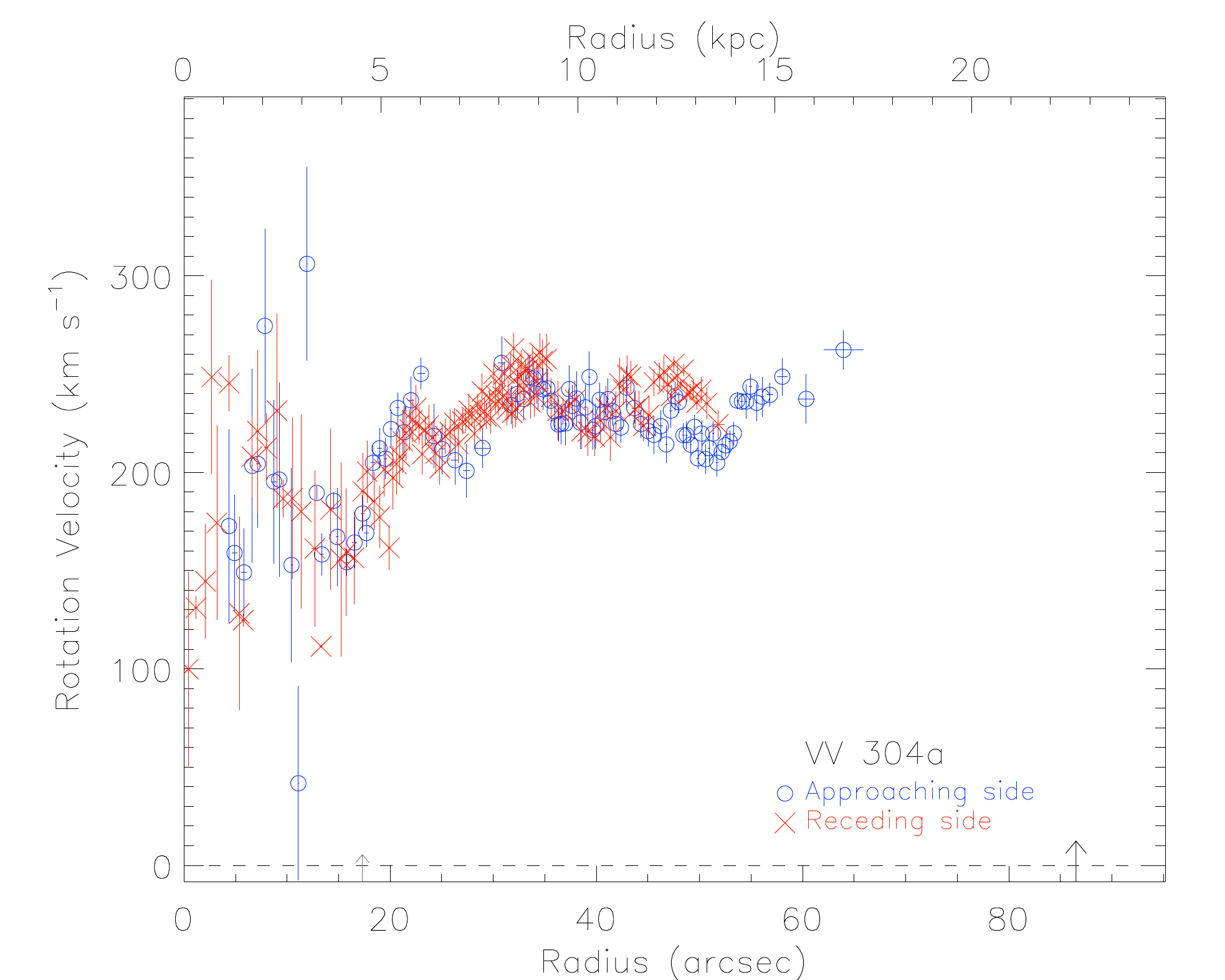}\\
\includegraphics[width=0.93\columnwidth]{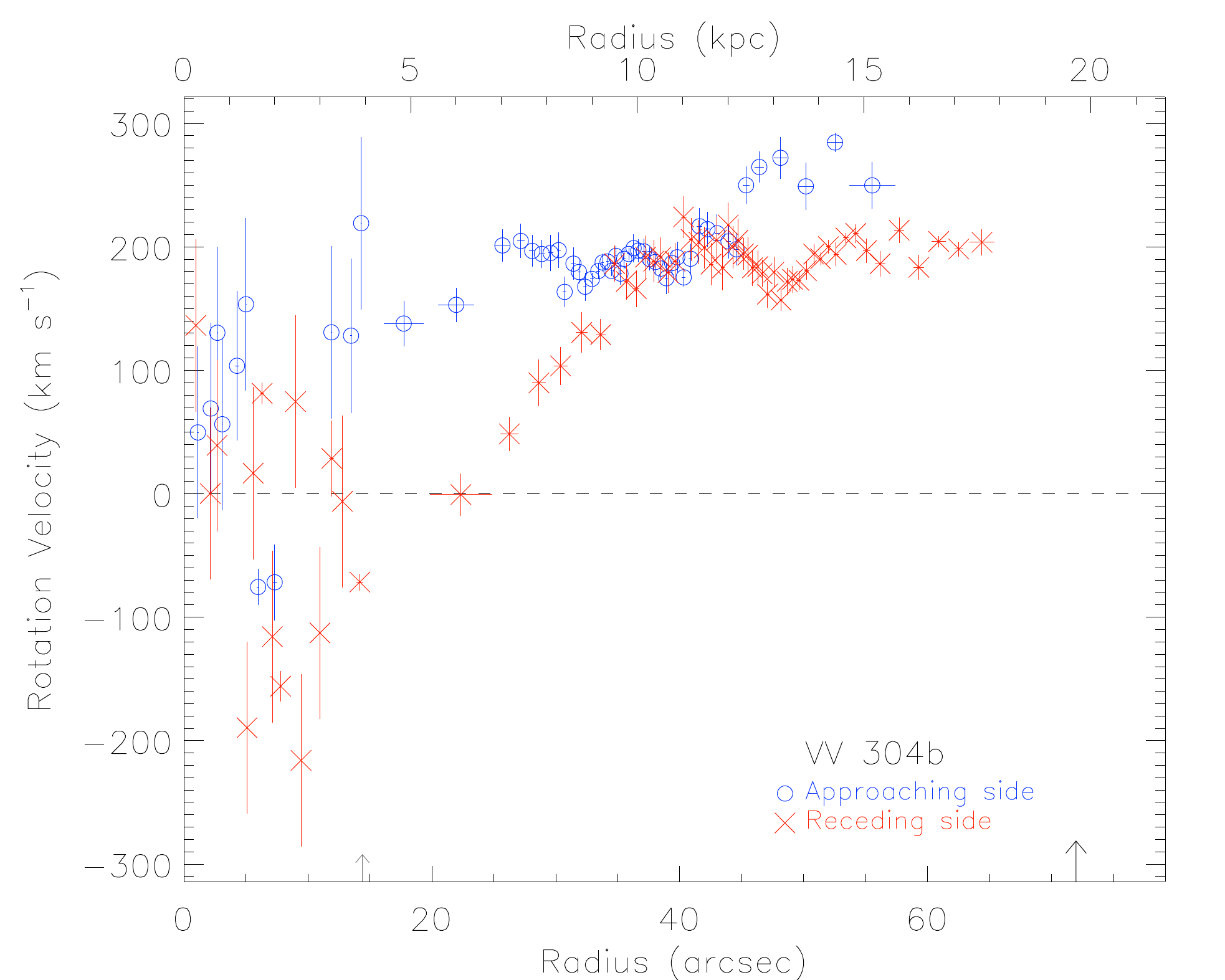}
\includegraphics[width=0.93\columnwidth]{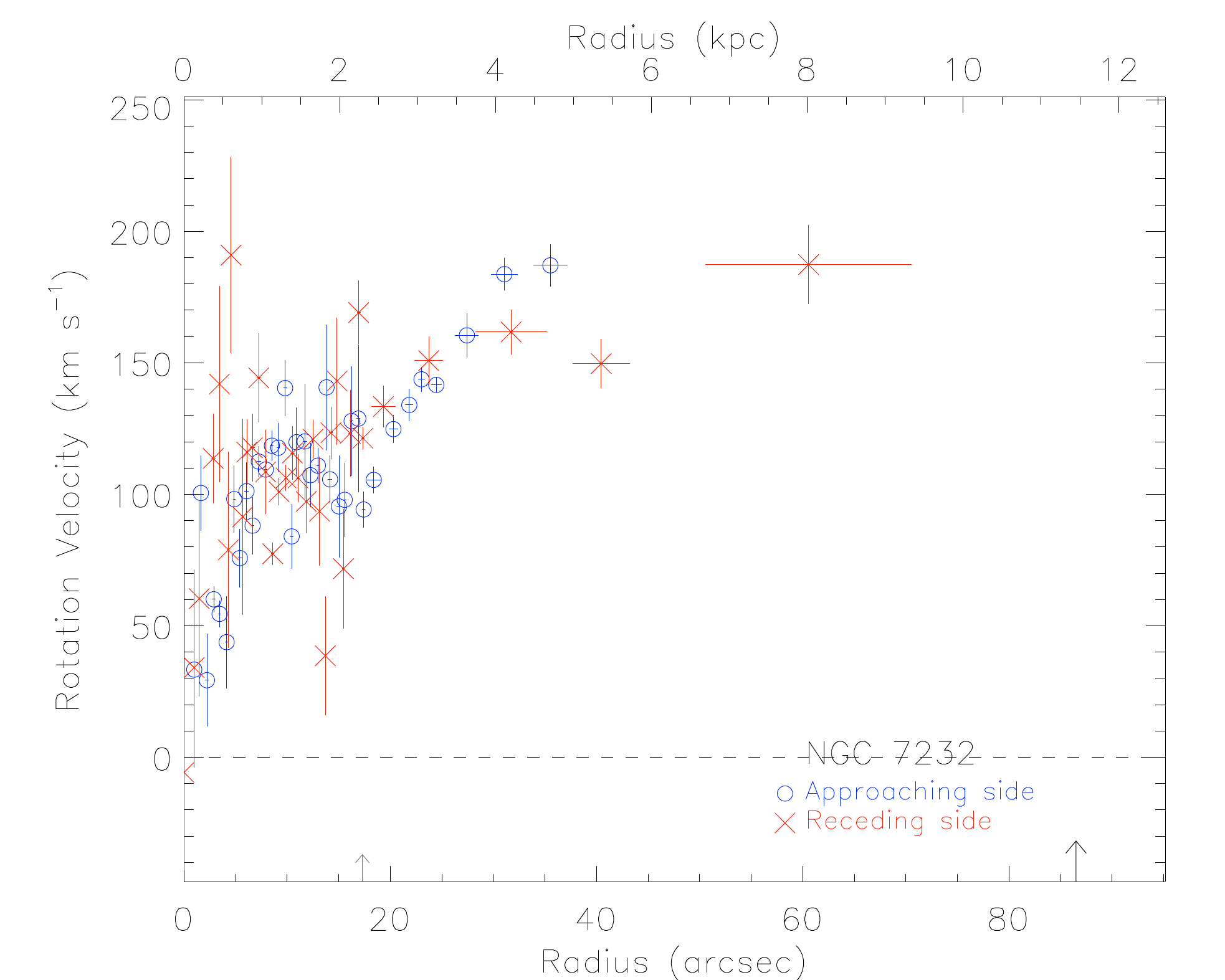}\\
\includegraphics[width=0.93\columnwidth]{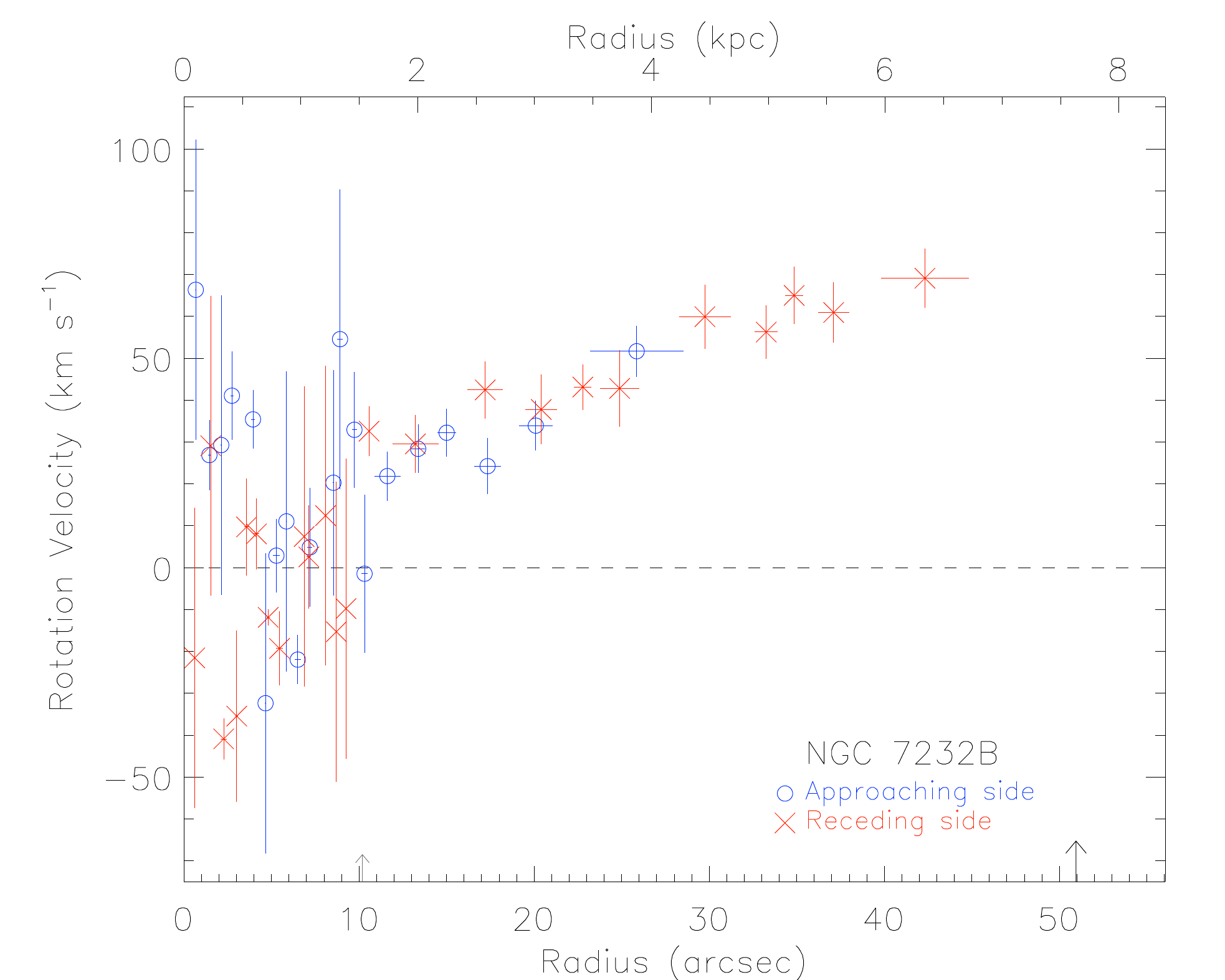}
\includegraphics[width=0.93\columnwidth]{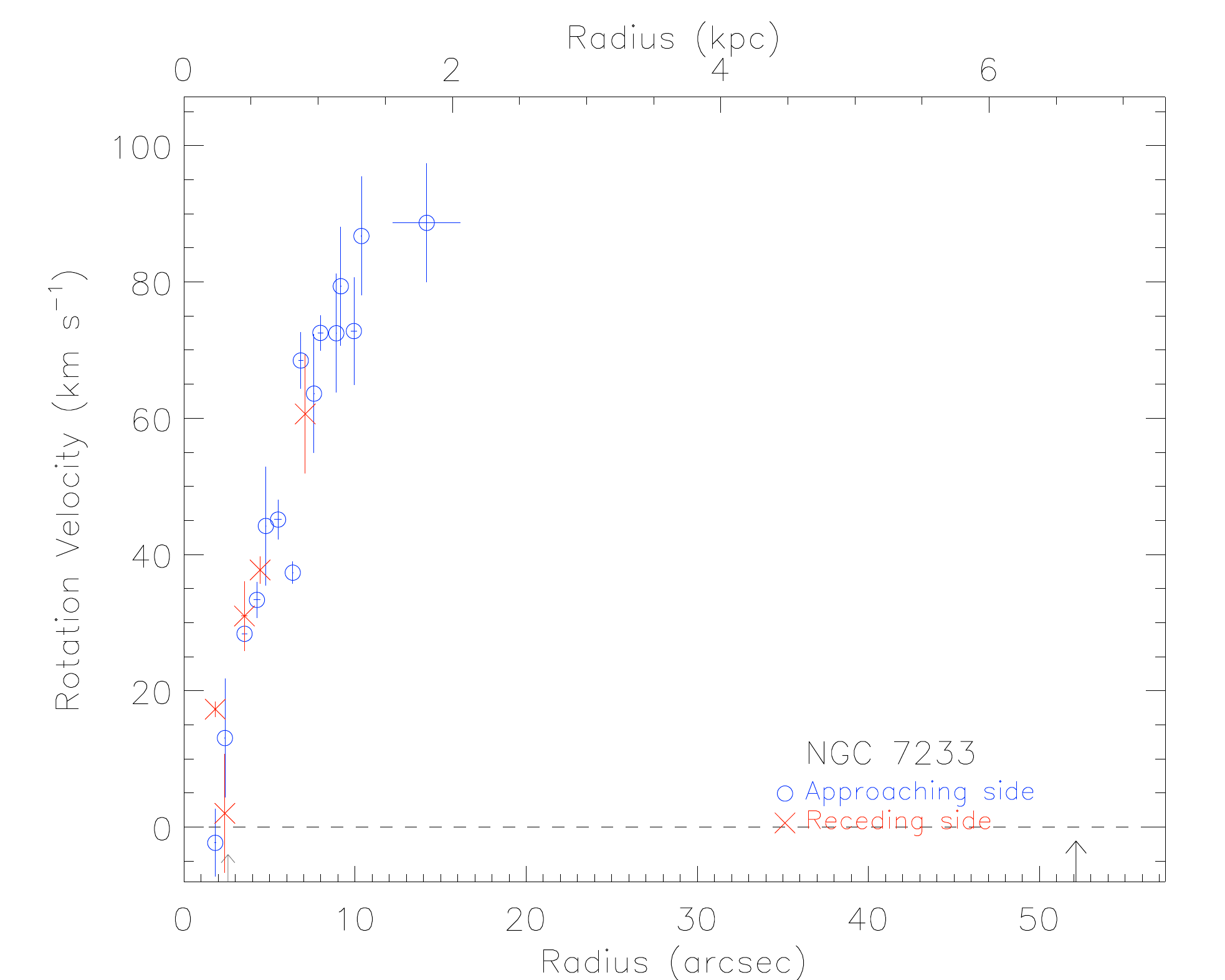}
\caption{Continued}
\label{rc2}
\end{figure*}

\begin{figure*}
\centering
\ContinuedFloat
\includegraphics[width=0.93\columnwidth]{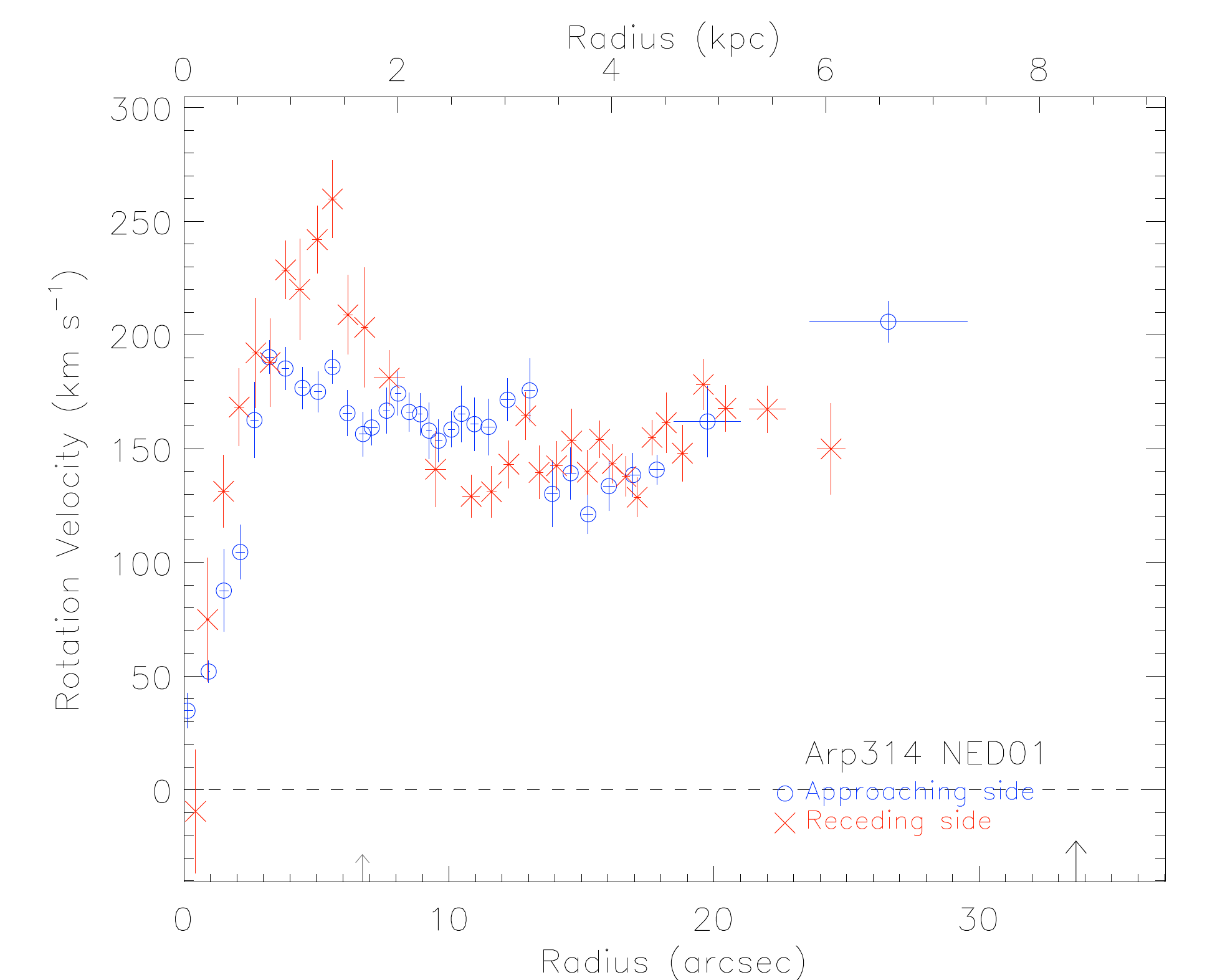}
\includegraphics[width=0.93\columnwidth]{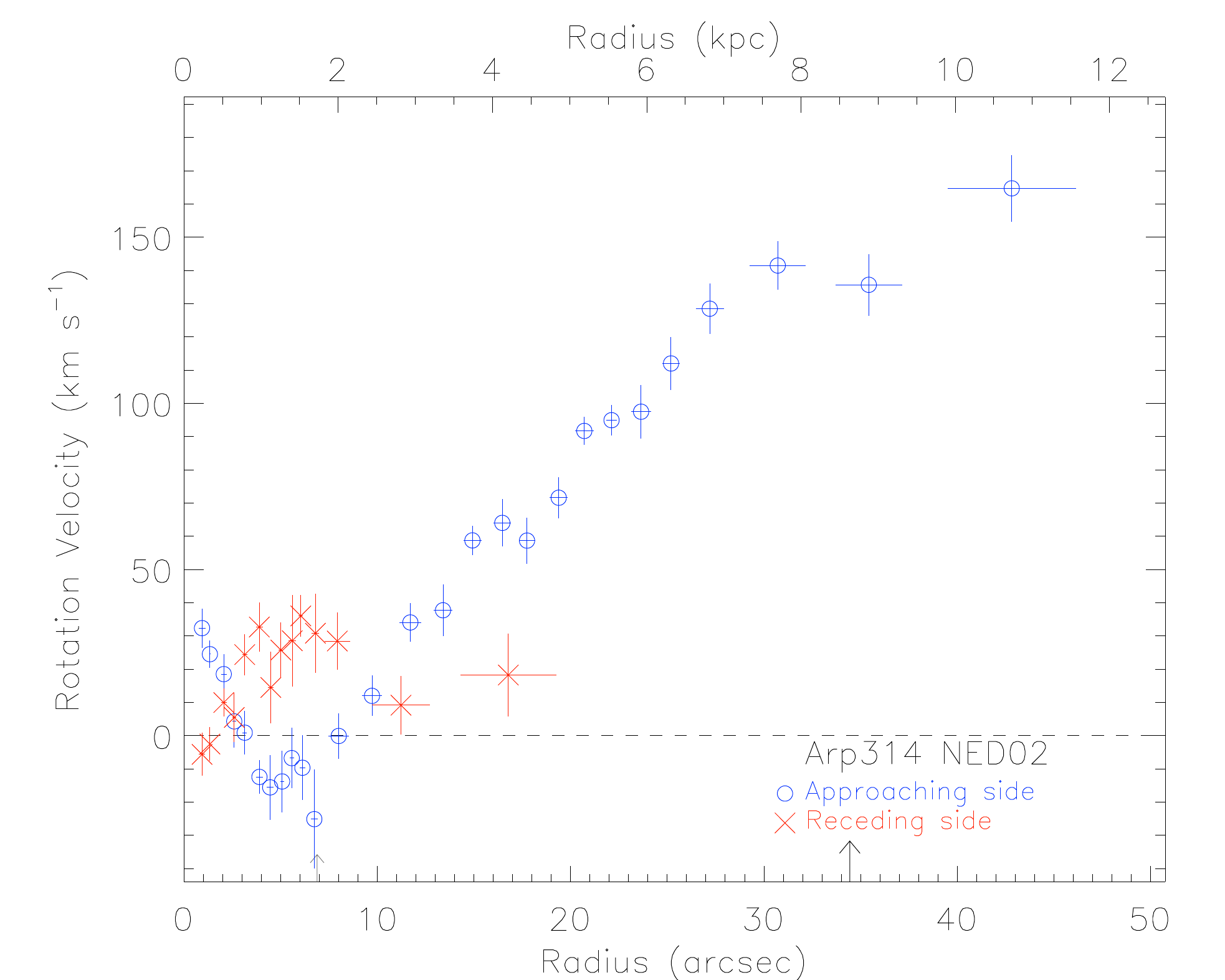}
\includegraphics[width=0.93\columnwidth]{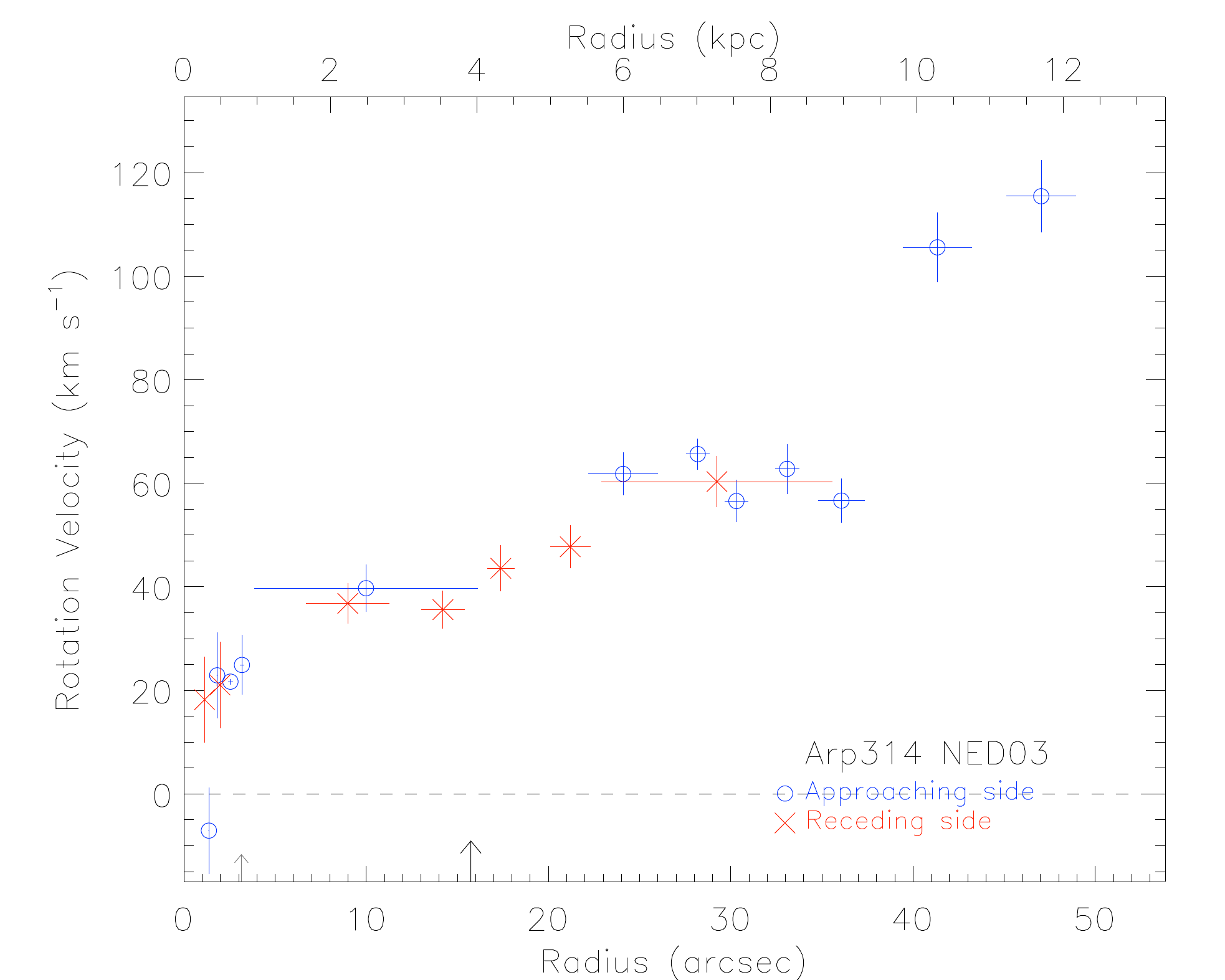}\\
\caption{Continued}
\label{rc3}
\end{figure*}

\label{lastpage}
\end{document}